\documentclass[pdftex,twocolumn,epjc3]{svjour3}          

\RequirePackage[T1]{fontenc}

\smartqed  

\RequirePackage{graphicx}
\RequirePackage{mathptmx}      
\RequirePackage{flushend}
\RequirePackage[numbers,sort&compress]{natbib}
\RequirePackage[colorlinks,citecolor=blue,urlcolor=blue,linkcolor=blue]{hyperref}

\journalname{Eur. Phys. J. C}

\usepackage{amsmath}
\usepackage{float}
\usepackage{subfigure}
\usepackage[colorlinks,linkcolor=red]{}

\usepackage{tikz,xcolor}
\usepackage{soul}
\sethlcolor{yellow}
\soulregister\texttt7
\newcommand{\revised}[1]{#1}
\newcommand{\mathrevised}[1]{\ensuremath{#1}}

\hypersetup{
  colorlinks=true,
  linkcolor=blue,
  urlcolor=blue,
  citecolor=blue,
  linktoc=all
}

\definecolor{lime}{HTML}{A6CE39}
\DeclareRobustCommand{\orcidicon}{%
	\begin{tikzpicture}
	\draw[lime, fill=lime] (0,0) 
	circle [radius=0.16] 
	node[white] {{\fontfamily{qag}\selectfont \tiny ID}};
	\draw[white, fill=white] (-0.0625,0.095) 
	circle [radius=0.007];
	\end{tikzpicture}
	\hspace{-2mm}
}

\foreach \x in {A, ..., Z}{%
	\expandafter\xdef\csname orcid\x\endcsname{\noexpand\href{https://orcid.org/\csname orcidauthor\x\endcsname}{\noexpand\orcidicon}}
}

\begin{document}

\sloppy

\title{Tests of general relativity using analytic derivatives of parametrized post-Einsteinian gravitational waveforms within the Fisher-matrix framework}

\author{Jie Wu\thanksref{addr1,addr2}\orcidA{}
        \and
        Mengfei Sun\thanksref{addr1,addr2}\orcidB{}
        \and
        Jin Li\thanksref{addr1,addr2,addr3,e} \orcidC{}
}
\thankstext{e}{e-mail: \href{cqujinli1983@cqu.edu.cn}{cqujinli1983@cqu.edu.cn}}

\institute{
College of Physics, Chongqing University, Chongqing 401331, China\label{addr1}
\and
Department of Physics and Chongqing Key Laboratory for Strongly Coupled Physics, Chongqing University, Chongqing 401331, China\label{addr2}
 \and
Institute of  Advanced Interdisciplinary Studies, Chongqing University, Chongqing 401331, China\label{addr3}
}

\date{Received: date / Accepted: date}

\maketitle

\begin{abstract}
Testing gravity beyond general relativity (GR) is essential for probing fundamental physics in the strong-field and highly dynamical regime accessed by gravitational-wave (GW) observations.
In this work, we derive analytic expressions for waveform derivatives in the Fisher-matrix formalism within the parametrized post-Einsteinian framework, using the frequency-domain inspiral waveform.
These analytic derivatives enable stable and efficient Fisher-matrix calculations without relying on finite-difference schemes.
We apply this method to a wide range of detector configurations, including space-based, ground-based, and multiband observations, and combine it with different binary black hole population models.
Our results reveal clear and systematic trends in the constraints on non-GR effects as functions of post-Newtonian order, detector type, and source population.
They also demonstrate the complementarity between space- and ground-based detectors, particularly for effects that accumulate during the low-frequency inspiral.
The analytic approach substantially reduces computational cost and avoids numerical systematics associated with step-size choices, making it well suited for large-scale parameter studies.
These results provide robust forecasts for the capability of future GW observations to constrain a broad class of non-GR effects and environmental influences, highlighting the scientific potential of upcoming detector networks for precision tests of gravity.
\end{abstract}

\section{Introduction}
Since the formulation of general relativity (GR), its predictions have been tested extensively through Solar System experiments and astrophysical observations~\cite{test_GR,test_GR1,test_GR2}.
These tests have confirmed the validity of GR in the weak field regime with high precision, establishing it as the standard theory of gravitation~\cite{test_GR3,test_GR4}.
Despite these successes, several open issues remain.
In particular, the phenomena commonly attributed to dark matter and dark energy are not naturally explained within GR and have motivated continued interest in modified theories of gravity~\cite{test_GR5,test_GR6}.
Moreover, direct tests of gravity in the strong field and highly nonlinear regime remain comparatively scarce~\cite{test_GR7}.
Many potential deviations from GR are expected to manifest predominantly in strong gravitational fields, where nonlinear dynamics is essential and cannot be fully probed by weak field experiments alone~\cite{strong_field}.

From both theoretical and observational perspectives, extending tests of gravity into the strong field regime is therefore of central importance.
The direct detection of gravitational waves (GWs) has opened a new observational window for such tests~\cite{GW150914}.
GWs emitted by binary black hole (BBH) systems are generated by highly dynamical processes in the strong gravity regime and thus provide a unique probe of gravity under extreme conditions~\cite{CBC}.
With the detection of numerous BBH events by ground-based detectors, GW-based tests of GR have been carried out along several complementary directions~\cite{test_GR_with_BBH}.
These include inspiral merger ringdown consistency tests, residual tests, black hole (BH) spectroscopy, and so on~\cite{IMR_test,residual_test,no-hair}.
Applied to current observations, these methods have so far confirmed GR within observational uncertainties while placing bounds on possible deviations~\cite{GW250114,GWTC3_test}.

At present, ground-based detectors such as LIGO, Virgo, and KAGRA are operating and have reported more than two hundred GW events, most of which arise from mergers of stellar mass BBH (SBBH) systems~\cite{GWTC4}.
Future third generation detectors, such as ET, are expected to achieve significantly improved sensitivity.
This will allow the detection of a larger population of more distant stellar mass BBHs with longer observable inspiral signals~\cite{ET}.
Such observations will further enhance the precision of GW based tests of GR.
In parallel, space-based detectors including LISA, Taiji, and TianQin are designed to operate in the mHz frequency band~\cite{LISA,Taiji,TianQin}.
With much longer observation times, these missions will track the early inspiral of SBBHs and observe massive black hole binaries (MBHBs)~\cite{networks}.
They therefore provide complementary opportunities to test gravity in regimes inaccessible to ground-based detectors.

Beyond direct consistency checks between GW observations and GR predictions, an alternative approach is to parameterize possible deviations at the waveform level and constrain them using data~\cite{Parametrized_test}.
A representative example is the parametrized post-Einsteinian (ppE) framework, which introduces systematic corrections to the GW amplitude and phase relative to the GR waveform~\cite{ppE}.
Based on the post-Newtonian (PN) expansion, the ppE framework associates potential deviations with parameterized terms at specific PN orders.
This construction allows one to capture effects from a wide range of modified gravity scenarios without committing to a specific theory, while maintaining a clear connection to physical corrections in the waveform.
The ppE framework has been widely used in tests of GR with ground-based detectors~\cite{ppE_test_GR1,ppE_test_GR2,ppE_test_GR3,ppE_test_GR4}.
It provides a transparent and unified language for assessing the measurability of deviations from GR and serves as a natural basis for parameter estimation studies.

Within the ppE framework, a broad range of non-GR effects has been explored.
Tahura and Yagi derived ppE corrections to the GW amplitude and phase for several modified gravity theories and established a systematic mapping between physical mechanisms and their associated PN orders~\cite{ppE_both_A_Psi}.
Building on this work, Carson and Yagi used the Fisher matrix formalism to study constraints on parameterized phase corrections for different detector configurations and multiband observations~\cite{ppE_only_Psi}.
Focusing on specific observational scenarios, Liu \textit{et al.} combined sensitivity curves of space-based detectors to investigate the detectability of parameterized dipole radiation effects~\cite{Multiband_Observation}.
O’Beirne \textit{et al.} further extended the parameterized approach by including both propagation and radiation effects in a unified analysis~\cite{ppE_PTA}.
In the context of space-based detectors, parameterized tests at different PN orders have also been examined.
Using TianQin as an example, Shi \textit{et al.} studied how the constraining power depends on the PN order of the corrections~\cite{ppE_TianQin}.
The ppE framework has also been widely applied to studies of additional GW polarization modes, including analyses based on Taiji and the LISA-Taiji network~\cite{ppE_Taiji,ppE_LISA_Taiji}, as well as joint ground- and space-based observations of BBH signals~\cite{my_paper_ppE}.
These works highlight the central role of the ppE framework in model independent tests of gravity with GWs.
However, the above Fisher matrix analyses rely on numerical or semi-analytic waveform derivatives, which limits a systematic understanding of the dependence on PN order and the associated parameter correlations.

Motivated by these limitations, we develop an analytic Fisher matrix framework within the ppE formalism.
Using the \texttt{TaylorF2} waveform, we construct fully analytic derivatives of the inspiral ppE waveform with respect to all binary parameters and ppE parameters, consistently including corrections to both the amplitude and the phase.
Based on this framework, we perform a systematic study of the constraints on non-GR effects, with particular attention to their dependence on PN order, binary properties, and detector configurations.
We further consider several representative modified gravity theories and dynamical friction effects, for which the corresponding analytic waveform derivatives are derived explicitly.
Constraints are evaluated under different population models and detector networks, including multiband observation scenarios.
This analytic approach avoids numerical instabilities associated with finite difference derivatives and allows for a clearer interpretation of scaling behaviors and parameter correlations.
It therefore provides a transparent and controlled tool for model independent tests of gravity with GW observations.

This paper is organized as follows.
In Sec.~\ref{sec:waveform}, we introduce the waveform model and summarize the ppE framework adopted in this work.
Section~\ref{sec:Fisher_matrix} presents the derivation of the analytic waveform derivatives used in the Fisher matrix analysis.
Detector configurations, BBH population models, and parameter settings are described in Sec.~\ref{sec:Detectors_and_population}.
In Sec.~\ref{sec:Constraints_on_PN}, we examine how different PN coefficients and parameters affect the resulting constraints in the full parameter space.
Section~\ref{sec:Constraints_on_Specific} focuses on several representative non-GR effects and compares the constraining power of different detectors.
Finally, Sec.~\ref{sec:Conclusions} summarizes our main results.
\ref{app:PN_phase} provides explicit PN phase expressions and their derivatives, while \ref{app:ppE} lists the non-GR effects considered in this work together with their analytic waveform derivatives.
Throughout this paper, we use geometric units with $c=G=1$.

\section{Parametrized post-Einsteinian gravitational waveforms}\label{sec:waveform}
GW signals from compact binary systems are commonly divided into three stages.
These are the inspiral, merger, and ringdown phases.
Among them, the inspiral phase lasts the longest and can be accurately modeled by expanding the orbital energy and GW radiation flux in a PN series~\cite{PN_Theory}.
In this regime, a frequency domain representation of the signal is particularly convenient.
The most widely used approximation is based on the stationary phase approximation (SPA).
\revised{In this work, we adopt the frequency domain SPA waveform \texttt{TaylorF2} as the aligned-spin, quasi-circular PN waveform template in GR}~\cite{TaylorF2}.
It can be written as
\begin{equation}
	\tilde h_{\rm GR}(f)=\mathcal A f^{-7/6} e^{i\Psi_{\rm GR}(f)} .
\end{equation}
The amplitude $\mathcal A$ is given by~\cite{PN_amp}
\begin{equation}
	\mathcal A=\sqrt{\frac{5}{24}} \pi^{-2/3}
	\frac{\mathcal M^{5/6}}{D_L} \mathcal C ,
\end{equation}
where $\mathcal M=(m_1 m_2)^{3/5}/(m_1+m_2)^{1/5}$ is the chirp mass, $m_1$ and $m_2$ denote the component masses of the BBH, and $D_L$ is the luminosity distance.
The factor $\mathcal C$ encodes the detector response to the source sky location and polarization.
Since the detector sensitivity curves used in this work are averaged over sky position and polarization, we set $\mathcal C=1$ to isolate the effects of PN and ppE corrections~\cite{LISA_sensitivity}.

The phase $\Psi_{\rm GR}(f)$ is expressed as a PN expansion~\cite{PN_coefficients}
\begin{equation}
	\Psi_{\rm GR}(f)
	= 2\pi f t_c - \phi_c - \frac{\pi}{4}
	+ \frac{3}{128\eta} \sum_{p=0}^{4} \varphi_p v^{p-5} .
\end{equation}
The PN expansion parameter is defined as
\begin{equation}
	v=(\pi f M)^{1/3} ,
\end{equation}
where $t_c$ and $\phi_c$ denote the coalescence time and phase, $M=m_1+m_2$ is the total mass, and $\eta=m_1 m_2/M^2$ is the symmetric mass ratio. 
Explicit expressions for the PN coefficients $\varphi_p$ are summarized in \ref{app:PN_phase}.

\revised{In this work, the GR PN expansion is truncated at 2PN order.}
This choice is motivated by two considerations.
First, most ppE corrections studied in the literature enter at PN orders no higher than 2PN.
The adopted truncation therefore captures the dominant non-GR effects~\cite{ppE_both_A_Psi}.
\revised{Second, a related assessment examined how different PN waveform models affect parameter estimation for SBBHs observed by space-based GW detectors}~\cite{my_paper_PN}.
\revised{That study differs from the present work, which focuses on how ppE corrections at different PN orders affect the constraints on non-GR effects, but it provides a useful reference for choosing the baseline GR waveform.}
\revised{In particular, the 2PN inspiral waveform is sufficient for the Fisher-matrix forecast and scaling analysis performed here.}
Higher order PN corrections primarily modify the phase evolution.
They can be incorporated straightforwardly by extending the PN phase expansion.
Such extensions do not affect the overall structure of the analytic framework developed in this work.

We note that more complete waveform models are available, including models that incorporate eccentricity or spin-precession effects~\cite{TaylorF2Ecck,SEOBNREP}.
These models are essential when the corresponding physical effects are part of the parameter space to be inferred.
However, the present analysis does not aim to measure additional source parameters such as the eccentricity or spin tilt angles.
For the parameters considered here, the systematic biases arising from neglecting such effects are expected to be subdominant to the statistical uncertainty within the regime of validity of our approximation~\cite{systematic_uncertainties}.
This expectation is also consistent with the fact that isolated BBHs are expected to have negligible eccentricity, while a large fraction of dynamically formed BBHs are also expected to retain only small eccentricities~\cite{dynamically_formed_BBH}.
Therefore, the use of the aligned-spin, quasi-circular \texttt{TaylorF2} approximant represents a practical compromise between computational efficiency and waveform accuracy for the purpose of the present Fisher-matrix study.
A related assessment is presented in our previous work~\cite{my_paper_PN}.

The PN waveform introduced above provides an accurate description of inspiral signals within GR.
In more general theories of gravity, however, modifications to the orbital dynamics or radiation mechanisms may introduce systematic deviations in both the waveform amplitude and phase.
To capture such effects without committing to a specific theory, a model independent parameterization is required.
Because the inspiral waveform exhibits a well defined PN power law structure, the minimal extension of the GR waveform can be formulated as power law corrections in the frequency domain.
If non-GR effects enter at specific PN orders, or if only the leading contribution is retained, these deviations can be incorporated consistently through the ppE framework.
This approach provides a unified parameterization of possible departures from GR.
Within the ppE framework, the frequency domain waveform takes the form~\cite{ppE}
\begin{equation}
	\tilde h(f)=\tilde h_{\rm GR}(f)\left(1+\alpha u^a\right)e^{i\beta u^b},
\end{equation}
with
\begin{equation}
	u=(\pi f \mathcal M)^{1/3},
\end{equation}
where $\alpha$ and $\beta$ quantify relative corrections to the amplitude and phase with respect to the GR waveform.
The GR limit is recovered when $\alpha=\beta=0$.
The exponents $a$ and $b$ determine the PN order at which the corrections enter.
They are related to the PN counting through
\begin{equation}
	\mathrevised{a = 2 n_{\rm PN}, \qquad b = 2 n_{\rm PN}-5 .}
\end{equation}
\revised{Here} \mathrevised{n_{\rm PN}} \revised{denotes the PN order of the ppE correction term.} \revised{In our analysis, the standard GR waveform is kept at 2PN order, while the ppE correction is assigned a PN order} \mathrevised{n_{\rm PN}} \revised{and varied to study the PN-order dependence of the constraints.}
By varying the coefficients and exponents, the ppE framework captures the leading inspiral deviations predicted by a wide class of modified gravity theories.

In general, the ppE parameters $\alpha$ and $\beta$ are not independent constants.
They depend on both the underlying theory and the intrinsic parameters of the binary system.
Theory specific coupling constants, mass ratios, or spin parameters may contribute to the amplitude and phase corrections in characteristic ways.
To retain the unified ppE form while separating the overall deviation strength from its parametric dependence, we further reparameterize the correction function as
\begin{equation}
	\alpha = \delta \Xi(\theta_{\rm int}),\qquad 
	\beta = \delta \Upsilon(\theta_{\rm int}),
\end{equation}
where $\delta$ denotes the deformation parameter that characterizes the non-GR effect.
The functions $\Xi(\theta_{\rm int})$ and $\Upsilon(\theta_{\rm int})$ encode the dependence on the intrinsic parameters $\theta_{\rm int}$ of the BBH.
With this reparameterization, the waveform adopted in this work can be written in the unified form
\begin{equation}\label{eq:waveform}
	\tilde h(f)=\mathcal A \mathcal R f^{-7/6}
	e^{i\left[\Psi_{\rm GR}(f)+\Psi_{\rm NGR}(f)\right]},
\end{equation}
where
\begin{equation}
\begin{aligned}
	&\mathcal R = 1+\delta \Xi(\theta_{\rm int}) u^a ,\\
	&\Psi_{\rm NGR} = \delta \Upsilon(\theta_{\rm int}) u^b .
\end{aligned}
\end{equation}
This formulation provides a direct starting point for analytic evaluations of waveform derivatives with respect to all parameters.
It is therefore well suited for the Fisher matrix analysis developed in the following sections.

In summary, this section has introduced the GW waveform model used in this work.
The inspiral signal is described using the frequency domain PN waveform \texttt{TaylorF2} within GR.
Possible non-GR effects are incorporated through the ppE framework, which provides a unified and flexible parameterization.
Building on this waveform model, the next section presents the Fisher matrix formalism and derives the required analytic waveform derivatives.

\section{Fisher matrix formalism with analytic waveform derivatives}\label{sec:Fisher_matrix}
In this work, the parameter set used in the Fisher-matrix analysis includes the intrinsic mass and spin parameters of the BBH system, the extrinsic parameters, and the ppE parameter characterizing non-GR effects.
It is given by
\begin{equation}\label{eq:theta}
	\theta=\{\ln\mathcal M,\ \ln\eta,\ \chi_s,\ \chi_a,\ \ln D_L,\ t_c,\ \phi_c,\ \delta\},
\end{equation}
\revised{where} \mathrevised{\mathcal M} \revised{is the chirp mass,} \mathrevised{\eta} \revised{is the symmetric mass ratio,} \mathrevised{D_L} \revised{is the luminosity distance,} \mathrevised{t_c} \revised{and} \mathrevised{\phi_c} \revised{are the coalescence time and phase, and} \mathrevised{\delta} \revised{is the deformation parameter.}
This choice captures the key physical degrees of freedom governing the inspiral waveform.
It also allows a unified treatment of correlations between the binary parameters and the non-GR deformation parameter within a single framework.
The symmetric and antisymmetric spin combinations are defined as
\begin{equation}
\chi_s =\frac{\chi_1+\chi_2}{2},\qquad
\chi_a = \frac{\chi_1-\chi_2}{2},
\end{equation}
where $\chi_{1,2}$ are the dimensionless spins of the two BHs.
These combinations naturally arise in the PN description of the inspiral phase.
They are therefore convenient variables for parameter estimation analyses.

For the parameter set defined above, the Fisher matrix is constructed as
\begin{equation}\label{eq:Fisher_matrix}
	\Gamma_{ij}
	= 4 \Re\int_{f_{\min}}^{f_{\max}}
	\frac{\partial_{\theta_i}\tilde h(f)\,
	\partial_{\theta_j}\tilde h^{*}(f)}
	{S_n(f)} \mathrm{d}f ,
\end{equation}
where $S_n(f)$ denotes the detector sensitivity curve.
The inverse of the Fisher matrix defines the covariance matrix of the parameters, whose diagonal elements correspond to the variances~\cite{FIM1,FIM2}.
Accordingly, the statistical uncertainty of a parameter $\theta$ is given by
\begin{equation}
	\sigma_{\theta}
	= \sqrt{\left(\Gamma^{-1}\right)_{\theta\theta}} .
\end{equation}
When multiple GW detectors are considered and their noises are statistically independent, the total Fisher matrix is obtained by summing the individual contributions.
If the Fisher matrix associated with the $n$th detector is denoted by $\Gamma^{(n)}_{ij}$, the Fisher matrix for the joint observation can be written as
\begin{equation}
	\Gamma_{ij}=\sum_{n}\Gamma^{(n)}_{ij}.
\end{equation}
Throughout this work, all results are obtained by combining the Fisher matrices of individual detectors or detector networks.
The resulting parameter uncertainties are evaluated from the inverse Fisher matrix.
This allows a direct comparison of the constraining power of different detectors and detector networks.

To derive the waveform partial derivatives required for the Fisher-matrix analysis in a unified form, we reorganize the waveform derivatives following the strategy of Ref.~\cite{Fisher_analytical}.
We first express the derivative of the waveform with respect to an arbitrary parameter $\theta$ in logarithmic form.
Using the exponential structure of the frequency-domain waveform, one can write
\begin{equation}
	\partial_{\theta}\tilde h
	=\tilde h \partial_{\theta}\ln\tilde h .
\end{equation}
Since the waveform adopted in this work contains both the GR contribution and the non-GR correction introduced through the ppE framework, substituting Eq.~(\ref{eq:waveform}) yields
\begin{equation}
	\partial_{\theta}\tilde h
	=\tilde h\left(
	\partial_{\theta}\ln \mathcal A
	+\partial_{\theta}\ln \mathcal R
	+i \partial_{\theta}\Psi_{\rm GR}
	+i \partial_{\theta}\Psi_{\rm NGR}
	\right).
\end{equation}
This expression shows that the influence of each parameter can be separated into contributions to the amplitude and to the phase. 
To simplify the notation, we define
\begin{equation}
\begin{aligned}
	X_\theta &= \partial_\theta\ln\mathcal A+\partial_\theta\ln\mathcal R,\\
	Y_\theta &= \partial_\theta\Psi_{\rm GR}+\partial_\theta\Psi_{\rm NGR}.
\end{aligned}
\end{equation}
The waveform derivative can then be written as
\begin{equation}
	\partial_{\theta}\tilde h=\tilde h (X_\theta+iY_\theta).
\end{equation}

With these definitions, the Fisher matrix can be expressed as
\begin{equation}\label{eq:Fisher_matrix_XY}
	\Gamma_{ij}
	=\int_{f_{\min}}^{f_{\max}}
	\frac{4|\tilde h|^2}{S_n(f)}
	\left(X_iX_j+Y_iY_j\right)\mathrm{d}f,
\end{equation}
where the squared magnitude of the waveform is given by
\begin{equation}
	|\tilde h|^2=\mathcal A^2 \mathcal R^2 f^{-7/3}.
\end{equation}
Within the same framework, the signal-to-noise ratio (SNR) can be written as
\begin{equation}
	{\rm SNR}^2=\int_{f_{\min}}^{f_{\max}}
	\frac{4|\tilde h|^2}{S_n(f)} \mathrm{d}f .
\end{equation}
This formulation provides a consistent basis for the analytic Fisher-matrix analysis presented in this work.

We compute the waveform derivatives with respect to all parameters and present explicit analytic expressions for $X_\theta$ and $Y_\theta$.
For parameters entering only the amplitude or only the phase, the derivatives take simple forms.
For $D_L$, $t_c$, and $\phi_c$, one finds
\begin{equation}
\begin{aligned}
X_{\ln D_L} = -1,\quad Y_{\ln D_L} = 0, \\
X_{t_c}    = 0, \quad Y_{t_c}    = 2\pi f, \\
X_{\phi_c} = 0, \quad Y_{\phi_c} = -1 .
\end{aligned}
\end{equation}
To treat amplitude and phase contributions uniformly for the remaining parameters, we introduce
\begin{equation}
	\kappa = \frac{\mathcal R-1}{\mathcal R}.
\end{equation}

The mass parameters affect both the amplitude normalization and the PN phase at multiple orders.
Including possible non-GR mass dependence, the derivatives are given by
\begin{equation}
\begin{aligned}
X_{\ln\mathcal M}
=\frac{5}{6}
+\kappa \left(
\frac{\partial\ln\Xi}{\partial\ln\mathcal M}
+\frac{a}{3}
\right), \\[0.5ex]
Y_{\ln\mathcal M}
=\frac{\partial\Psi_{\rm GR}}{\partial\ln\mathcal M}
+\Psi_{\rm NGR} \left(
\frac{\partial\ln\Upsilon}{\partial\ln\mathcal M}
+\frac{b}{3}
\right), \\[0.5ex]
X_{\ln\eta}
=\kappa 
\frac{\partial\ln\Xi}{\partial\ln\eta}, \\[0.5ex]
Y_{\ln\eta}
=\frac{\partial\Psi_{\rm GR}}{\partial\ln\eta}
+\Psi_{\rm NGR}
\frac{\partial\ln\Upsilon}{\partial\ln\eta}.
\end{aligned}
\end{equation}

Spin parameters contribute through higher-order PN terms and possible ppE corrections.
Their derivatives are
\begin{equation}
\begin{aligned}
X_{\chi_{s,a}}
=\kappa 
\frac{\partial\ln\Xi}{\partial\chi_{s,a}}, \\[0.5ex]
Y_{\chi_{s,a}}
=\frac{\partial\Psi_{\rm GR}}{\partial\chi_{s,a}}
+\Psi_{\rm NGR}
\frac{\partial\ln\Upsilon}{\partial\chi_{s,a}} .
\end{aligned}
\end{equation}

The deformation parameter $\delta$ enters only through non-GR corrections.
Its derivatives are therefore
\begin{equation}
	X_\delta=\frac{\Xi u^a}{1+\delta \Xi u^a},\qquad
	Y_\delta=\Upsilon u^b .
\end{equation}

All derivatives of the GR phase are listed in \ref{app:PN_phase}.
Given explicit forms of $\Xi$ and $\Upsilon$, the full set of waveform derivatives follows directly.
Representative non-GR cases are summarized in  \ref{app:ppE} and used in Sec.~\ref{sec:Constraints_on_Specific}.

\begin{figure}[ht]
    \begin{minipage}{\columnwidth}
        \centering
        \includegraphics[width=0.95\textwidth,
        trim=0 0 0 0,clip]{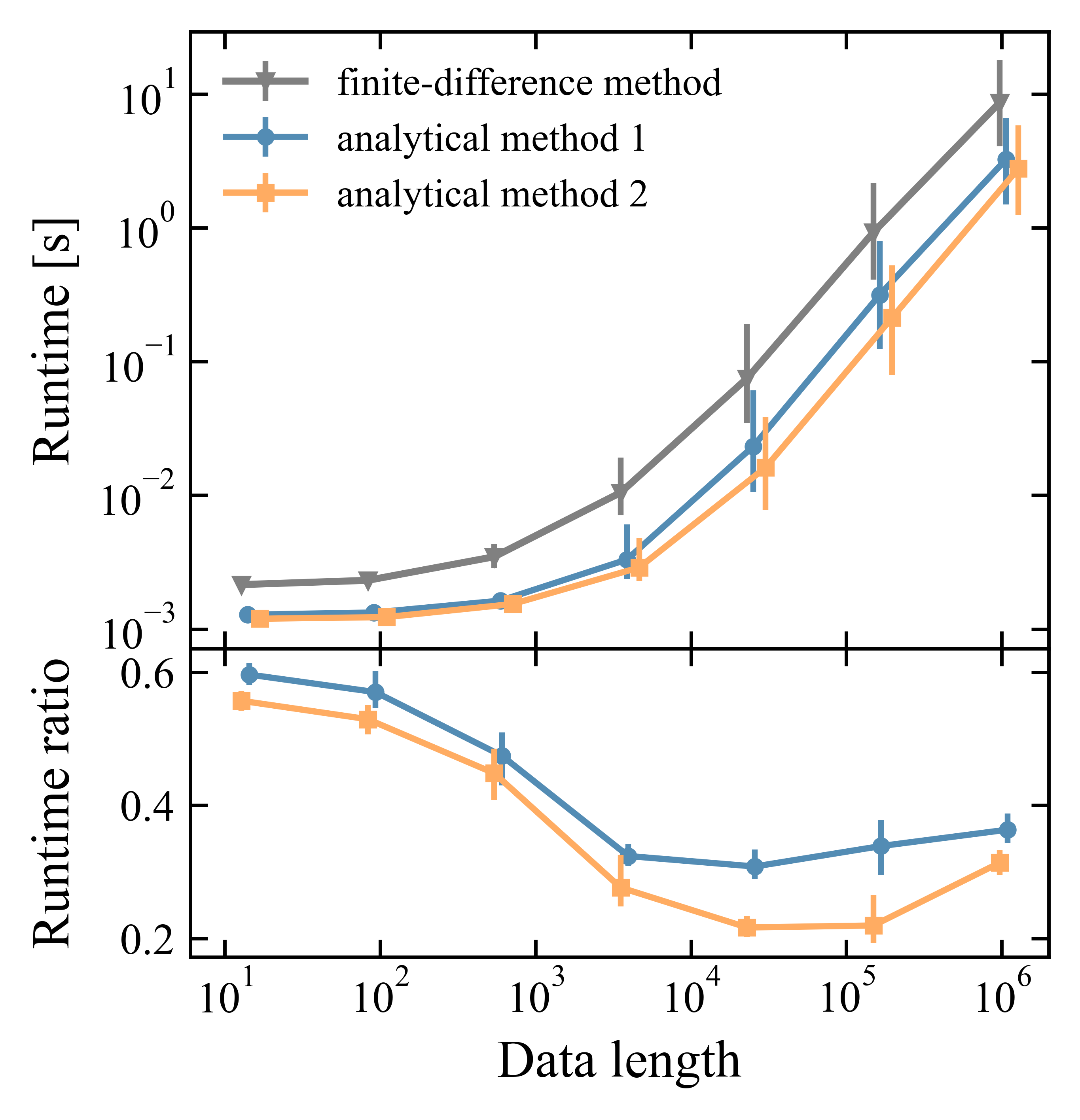}
        \caption{Comparison of the runtime for different methods. \revised{Data length denotes the length of the Python array, i.e., the number of sampled frequency points used in the runtime test.} The upper panel shows the absolute runtime of each method, while the lower panel displays the relative runtime normalized to the finite-difference method. Here, the \textit{finite-difference method} refers to the standard numerical evaluation of waveform derivatives. The analytic method denotes the use of our analytic derivatives: \textit{analytic method~1} constructs $\partial_{\theta}\tilde h=\tilde h (X_\theta+iY_\theta)$ and substitutes it into Eq.~(\ref{eq:Fisher_matrix}), whereas \textit{analytic method~2} directly inserts $X_\theta$ and $Y_\theta$ into Eq.~(\ref{eq:Fisher_matrix_XY}). Computations are performed on a dual Xeon Platinum 8480+ server with an NVIDIA A800 80 GB GPU.}\label{fig:Fisher_time}
    \end{minipage}
\end{figure}

Beyond providing a consistent and directly comparable derivation framework, analytic differentiation also offers clear advantages in numerical implementations. 
We compare the computational efficiency of different approaches, with the results shown in Fig.~\ref{fig:Fisher_time}. 
It shows that the \textit{analytic method} is significantly faster than the \textit{finite-difference method}, leading to a substantial reduction in computational cost. 
Moreover, \textit{analytic method~2} is faster than \textit{analytic method~1}, since it avoids the explicit evaluation of imaginary components in the calculation. 
In the most favorable cases, the runtimes of \textit{analytic method~1} and \textit{analytic method~2} are reduced to about $30.8\%$ and $21.6\%$ compared with \textit{finite-difference method}.
This improvement becomes critical when repeated Fisher-matrix evaluations are required.
Examples include large parameter scans and MCMC-based analyses.
In such cases, analytic derivatives substantially reduce the overall computational cost.

In summary, we construct a Fisher-matrix framework based on analytic waveform derivatives.
The amplitude-phase decomposition reduces the calculation to real-valued frequency integrals.
This framework provides the basis for all subsequent analyses.

\section{Detectors and population}\label{sec:Detectors_and_population}
\subsection{Space- and ground-based detectors}\label{subsec:detectors}
In the Fisher-matrix analysis, the role of a GW detector is fully encoded in its sensitivity curve and accessible frequency range.
For a fixed waveform model, detectors with different noise characteristics and operating bands respond differently to various frequency components of the signal.
This directly affects their ability to constrain physical parameters.
We therefore specify the detector configurations and noise models adopted in this work before presenting the results.

\begin{table}[ht]
\centering
\renewcommand{\arraystretch}{1.5}
\caption{Detector arm length and frequency band range.}\label{tab:detector}
\begin{tabular*}{\columnwidth}{@{\extracolsep{\fill}}lccc@{}}
\hline
 Detector & Arm length (km) & $f_{\rm low}$ (Hz) & $f_{\rm high}$ (Hz)\\
\hline
LISA & $2.5\times 10^6$ & $10^{-5}$ & $1$ \\
Taiji & $3\times 10^6$ & $10^{-5}$ & $1$ \\
TianQin & $\sqrt{3}\times 10^5$ & $10^{-4}$ & $1$ \\
\hline
LIGO & $4$ & $5$ & $5000$ \\
Virgo & $3$ & $10$ & $10000$ \\
KAGRA & $3$ & $1$ & $10000$ \\
ET & $10$ & $1$ & $10000$ \\
\hline
\end{tabular*}
\end{table}

We first consider space-based detectors.
Our analysis focuses on LISA, Taiji, and TianQin, which operate in the low-frequency band from the mHz to the Hz regime.
These detectors are particularly well suited for observing the long-lived inspiral signals emitted by BBH systems.
From a geometric perspective, LISA and Taiji employ heliocentric orbital configurations with arm lengths of order millions of kilometers.
In contrast, TianQin adopts a geocentric configuration with an arm length of order $10^5$ km.
These differences in arm length lead to distinct sensitivity shapes and low-frequency coverage.
As a result, different frequency intervals contribute with different weights in the Fisher-matrix integrals.

Assuming that other technical noise sources are effectively suppressed, the dominant noise contribution for space-based detectors arises from instrumental noise.
This includes optical path noise and test-mass acceleration noise.
The sensitivity curves used in this work are constructed from the corresponding noise power spectral densities, combined with the sky-averaged detector response.
Details of these constructions can be found in Refs.~\cite{LISA_sensitivity,Taiji_sensitivity,TianQin_sensitivity}.
In addition, we include the foreground noise produced by Galactic double white dwarf binaries.
This foreground may exceed the instrumental noise in the frequency range $\sim 0.5$--$3$ mHz and can affect the detectability of low-frequency signals.
A detailed discussion of both instrumental and foreground noise has been presented in our previous work~\cite{my_paper_confusion}.
The observation frequency ranges [$f_{\rm low},f_{\rm high}$] adopted for each space-based detector are summarized in Table~\ref{tab:detector}.

In this work, we do not consider laser frequency noise and therefore do not employ Time-Delay Interferometry (TDI). 
TDI is primarily used to suppress laser frequency noise, but it affects both GW signals and noise in the same way~\cite{Vallisneri2007}. 
Since we focus on statistical results rather than individual sources, the noise after TDI processing and the averaged response sensitivity curve are very close to the sensitivity curve used in our calculations. Therefore, they do not significantly change the overall results.
Our main goal is not to study the differences introduced by TDI. 
Considering the balance between computational cost and accuracy, we directly use the sensitivity curves in our calculations.
This approach is reasonable, and a more detailed comparison of TDI can be found in our previous work~\cite{my_paper_TDI}.

We next consider ground-based detectors.
Our analysis adopts the currently operating ground-based detector network LVK, including LIGO, Virgo, and KAGRA.
We also include the third-generation ground-based detector ET.
Ground-based detectors primarily operate in the high-frequency GW band above the Hz regime and are most sensitive to the late inspiral of BBH systems.
Their constraining capability is therefore complementary to that of space-based detectors.
The existing ground-based detectors have arm lengths at the kilometer scale.
LIGO employs 4 km arms, while Virgo and KAGRA each have arm lengths of 3 km.
For the third-generation detector ET, the effective arm length reaches the order of 10 km.
The increased arm length improves the low-frequency sensitivity, giving ET a clear advantage over current ground-based detectors in this regime.

The noise budget of ground-based detectors is more complex, as it includes multiple technical and environmental noise sources.
In the Fisher-matrix analysis presented here, we therefore directly adopt the design sensitivity curves of the detectors.
The sensitivity curve of LIGO is taken from LIGO Document T1800044.
Those of Virgo and KAGRA are taken from LIGO Document T1500293.
The sensitivity curve of ET is adopted from Ref.~\cite{ET_sensitivity}.
ET employs a triangular configuration.
When used in the Fisher-matrix analysis, the sensitivity of its equivalent L-shaped interferometer is multiplied by a geometric factor of 0.816.
The sensitivity curves of all ground-based detectors considered in this work are shown in Fig.~\ref{fig:population}(a).

\subsection{BBH selection and population model}\label{subsec:population}
\begin{figure*}[ht]
    \begin{minipage}{\textwidth}
        \centering
        \includegraphics[width=0.95\textwidth,
        trim=0 0 0 0,clip]{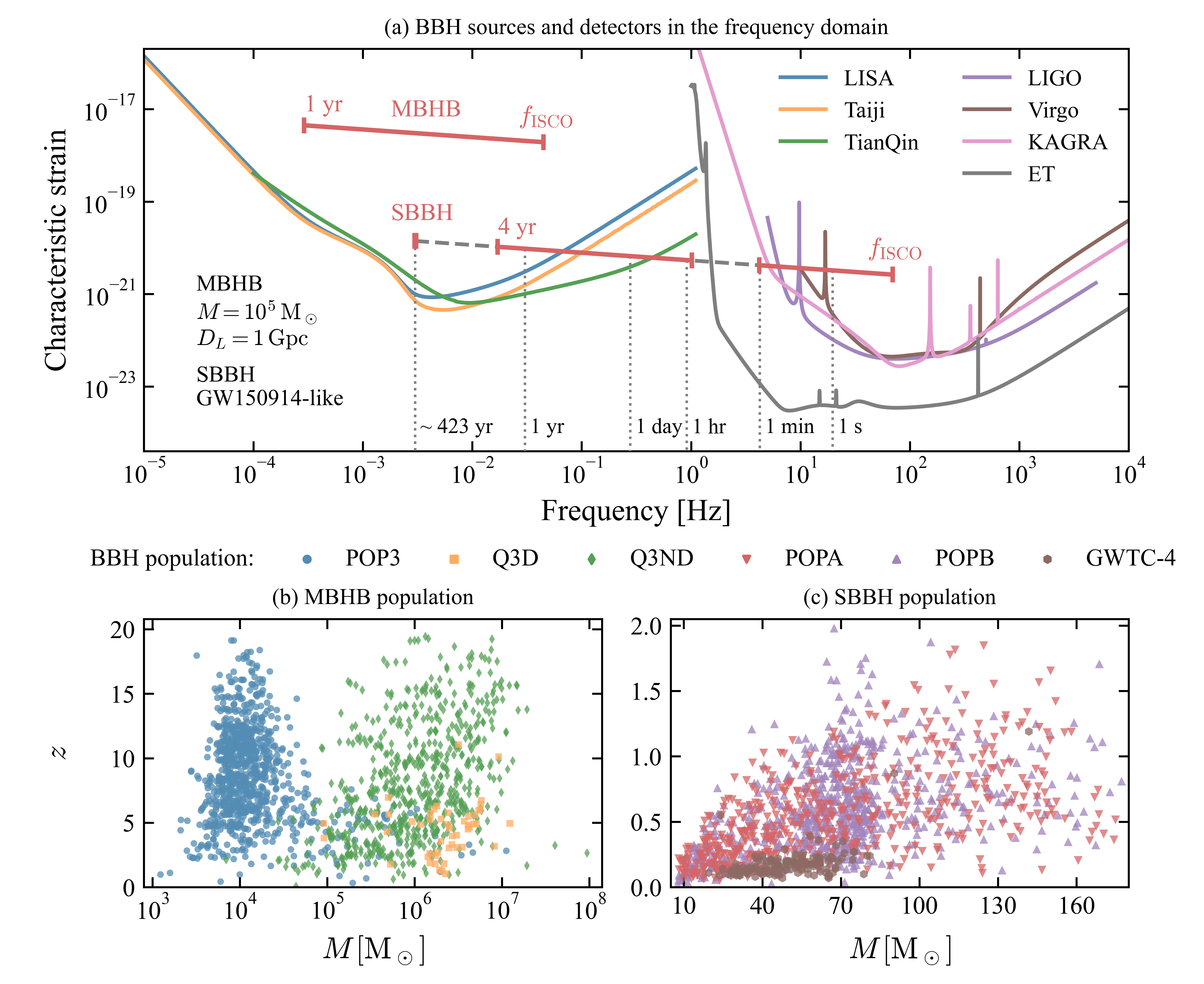}
        \caption{BBH selection and population models. (a) shows different BBH configurations in the frequency domain with the detector sensitivity curves. All times indicated in the figure correspond to the time to coalescence $\tau$. The red solid segments represent the portions of the BBH signals that are accessible to observations, while the dashed lines indicate the full inspiral evolution tracks. The detector sensitivity is shown in terms of the characteristic strain $\sqrt{f S_n(f)}$, whereas the BBH signals are plotted using the effective strain amplitude $2f|\tilde{h}(f)|$. (b) and (c) show the population models adopted for MBHBs and SBBHs, respectively. The MBHB population includes the \texttt{POP3}, \texttt{Q3D}, and \texttt{Q3ND} models, while the SBBH population consists of the \texttt{POPA}, \texttt{POPB}, and the latest LVK observational catalog \texttt{GWTC-4}~\cite{GWTC4}.}\label{fig:population}
    \end{minipage}
\end{figure*}

In the Fisher-matrix analysis, the mass scale and orbital evolution of a BBH system determine the frequency-domain distribution of the GW signal.
This directly affects the observable frequency range, the effective observation time, and the relative contribution of PN corrections at different orders.
Owing to their different operating frequency bands, space- and ground-based detectors exhibit complementary sensitivities to different GW sources.
Ground-based detectors can observe SBBH systems through the inspiral, merger, and ringdown phases.
Space-based detectors, by contrast, are capable of tracking the full coalescence of MBHB systems.
In addition, space-based detectors can observe the early inspiral of SBBHs at low frequencies, providing information on their long-term orbital evolution.

Because of the applicability limits of analytic calculations and PN waveform modeling, we restrict our analysis to the inspiral phase.
For a given BBH system, the upper limit of the frequency integration is taken to be the end of the inspiral prior to the innermost stable circular orbit (ISCO).
The corresponding frequency is given by~\cite{ISCO}
\begin{equation}
	f_{\mathrm{ISCO}}=\frac{1}{6\sqrt{6}\pi M}.
\end{equation}
The GW frequency emitted at a time $\tau$ before coalescence can be approximated as~\cite{PN_amp}
\begin{equation}\label{eq:f_tau}
	f(\tau)=\frac{1}{8\pi}\left(\frac{5}{\tau}\right)^{3/8}\mathcal{M}^{-5/8}.
\end{equation}
These relations imply that different detector configurations and BBH systems correspond to distinct observable frequency intervals and effective observation times.
Considering space-based detectors, ground-based detectors, and their multiband combinations, we classify the observational scenarios into four representative cases.
They are illustrated in Fig.~\ref{fig:population}(a).

\paragraph{space--MBHB:}
This case corresponds to high-SNR signals observed by space-based detectors.
We consider the inspiral of an MBHB system during an observation time $T_{\rm obs}$ prior to the ISCO, typically taking $T_{\rm obs}=1$ yr.
The corresponding frequency range is
\begin{equation}
	f_{\min}=f(\tau=T_{\rm obs}),\qquad 
	f_{\max}=f_{\mathrm{ISCO}}.
\end{equation}

\paragraph{space--SBBH:}
Space-based detectors observe only the early inspiral of SBBH systems, which are typically tens to hundreds of years away from merger.
Since such signals persist throughout the mission lifetime, we take $T_{\rm obs}=4$ yr.
The corresponding frequency range is approximated as
\begin{equation}
	f_{\min}=f(\tau),\qquad 
	f_{\max}=f(\tau-T_{\rm obs}).
\end{equation}

\paragraph{ground--SBBH:}
This case represents the primary source class for ground-based detectors.
We consider the inspiral of an SBBH system during an observation time $T_{\rm obs}$ prior to the ISCO, typically taking $T_{\rm obs}=10$ min.
The corresponding frequency range is
\begin{equation}
	f_{\min}=f(\tau=T_{\rm obs}),\qquad 
	f_{\max}=f_{\mathrm{ISCO}}.
\end{equation}

\paragraph{multiband--SBBH:}
This case describes SBBH systems observed first by space-based detectors and later by ground-based detectors within the mission duration.
The ground-based setup is identical to the ground--SBBH case.
For the space-based detector, the frequency range is taken to be
\begin{equation}
	f_{\min}=f(\tau=4 \mathrm{yr}),\qquad 
	f_{\max}=1 \mathrm{Hz}.
\end{equation}

In Fig.~\ref{fig:population}(a), we show representative frequency-domain trajectories for an MBHB signal and a GW150914-like SBBH signal.
The four red curves correspond to the four observational scenarios defined above.
The SBBH evolution is clearly separated into space-based, ground-based, and multiband regimes.

To assess parameter constraints under more realistic astrophysical conditions, we further introduce BBH population models.
Results based on a single source reflect the measurability at a specific point in parameter space.
In realistic observations, BBH systems span a wide range of masses, spins, and distances.

\begin{table}[ht]
    \centering
    \renewcommand{\arraystretch}{1.5}
    \caption{Parameter distribution used in calculation. $U[a,b]$ represents a uniform distribution from $a$ to $b$.}\label{tab:parameters}
    \begin{tabular*}{\columnwidth}{@{\extracolsep{\fill}}cc@{}}
    \hline
     Parameter & Distribution \\
    \hline
    lg($M$) [space] & $U[4,8]$  $\mathrm{M_\odot } $\\
    $M$ [ground] & $U[20,190]$  $\mathrm{M_\odot } $\\
    $q$ & $U[0.2,1]$ \\
    lg($z$) [space] & $U[-0.7, 1]$ \\
    lg($z$) [ground] & $U[-2.6, -0.7]$ \\
    $t_c$ & $U[0,100]$ min\\
    $\chi_{1,2}$ & $U[-0.95, 0.95]$ \\
    $\phi_c$ & $U[0,2\pi]$ rad\\
    $\tau$ & $U[5,500]/(1+z)$ yr\\
    \hline
    \end{tabular*}
\end{table}

We therefore evaluate the Fisher matrix over the full parameter space and incorporate physically motivated population models.
The parameter ranges and sampling strategies adopted in this work are summarized in Table~\ref{tab:parameters}.
For SBBHs observed by space-based detectors, the initial frequency is determined from Eq.~(\ref{eq:f_tau}) through $\tau$.
The luminosity distance $D_L$ is related to the redshift $z$ by the standard cosmological relation~\cite{waveform}
\begin{equation}
	D_L=\frac{1+z}{H_0}
	\int_0^z
	\frac{\mathrm{d}z'}{\sqrt{\Omega_m(1+z')^3+\Omega_\Lambda}} .
\end{equation}
We adopt the standard $\Lambda$CDM model with parameters from \textit{Planck 2018}~\cite{Planck_2018}, corresponding to the Hubble constant $H_0=67.37\ \mathrm{km}\ \mathrm{s}^{-1} \mathrm{Mpc}^{-1}$, matter density parameter $\Omega_m =0.315 $, and dark energy density parameter $\Omega_\Lambda  =0.685$. This full-parameter-space Fisher analysis is used in Sec.~\ref{sec:Constraints_on_PN} to study the dependence of parameter constraints on different PN-order corrections.
Building on this setup, we further introduce explicit BBH population models to characterize statistical properties of different source classes.
Representative examples are shown in Fig.~\ref{fig:population}(b).
For SBBHs observed by ground-based detectors, existing LVK observations already provide strong constraints.
Differences among astrophysical population models are therefore relatively small~\cite{LIGO_population1,LIGO_population2}.
We adopt a power-law mass distribution and consider two representative models.
The \texttt{popA} model excludes a peak component, while the \texttt{popB} model includes such a peak.
We also include the latest LVK event catalog \texttt{GWTC-4} as a real population model.

For MBHBs observed by space-based detectors, direct GW constraints are still limited.
We consider three representative models, \texttt{pop3}, \texttt{Q3d}, and \texttt{Q3nd}~\cite{TianQin_MBHB}.
The \texttt{pop3} model corresponds to the light-seed scenario.
The \texttt{Q3} models represent heavy-seed scenarios, with \texttt{Q3d} including a merger delay and \texttt{Q3nd} neglecting it.

For all six population models, source parameters are generated using \texttt{GWToolbox}~\cite{GWToolbox1,GWToolbox2}.
The corresponding parameter uncertainties are computed within the same Fisher-matrix framework.
In Sec.~\ref{sec:Constraints_on_Specific}, these population models are combined with specific non-GR theories to analyze detector-dependent constraints.

\section{Dependence of Constraints on PN Order and Binary Properties}\label{sec:Constraints_on_PN}
\subsection{Dependence on PN Order}
In this subsection, we begin with a general setup to study the dependence of parameter constraints on the PN order of the ppE correction.
For a broad class of modified gravity theories, the leading ppE corrections to the GW amplitude and phase can be written as power laws of $\eta$ \revised{with index} \mathrevised{n/5}\revised{.}
We assume
\begin{equation}
	\Xi(\eta)=\eta^{n/5},\qquad 
	\Upsilon(\eta)=\eta^{n/5}.
\end{equation}
\revised{In this notation,} \mathrevised{n} \revised{denotes the exponent parameter in} \mathrevised{\eta^{n/5}}\revised{, whereas} \mathrevised{n_{\rm PN}} \revised{denotes the PN order of the ppE correction.}
The coupling between non-GR effects and other system parameters is absorbed into the deformation parameter $\delta$.
This form has been shown to capture the inspiral corrections of many representative modified gravity theories~\cite{ppE_simple}.

\begin{figure}[ht]
    \begin{minipage}{\columnwidth}
        \centering
        \includegraphics[width=0.94\textwidth,
        trim=0 0 0 0,clip]{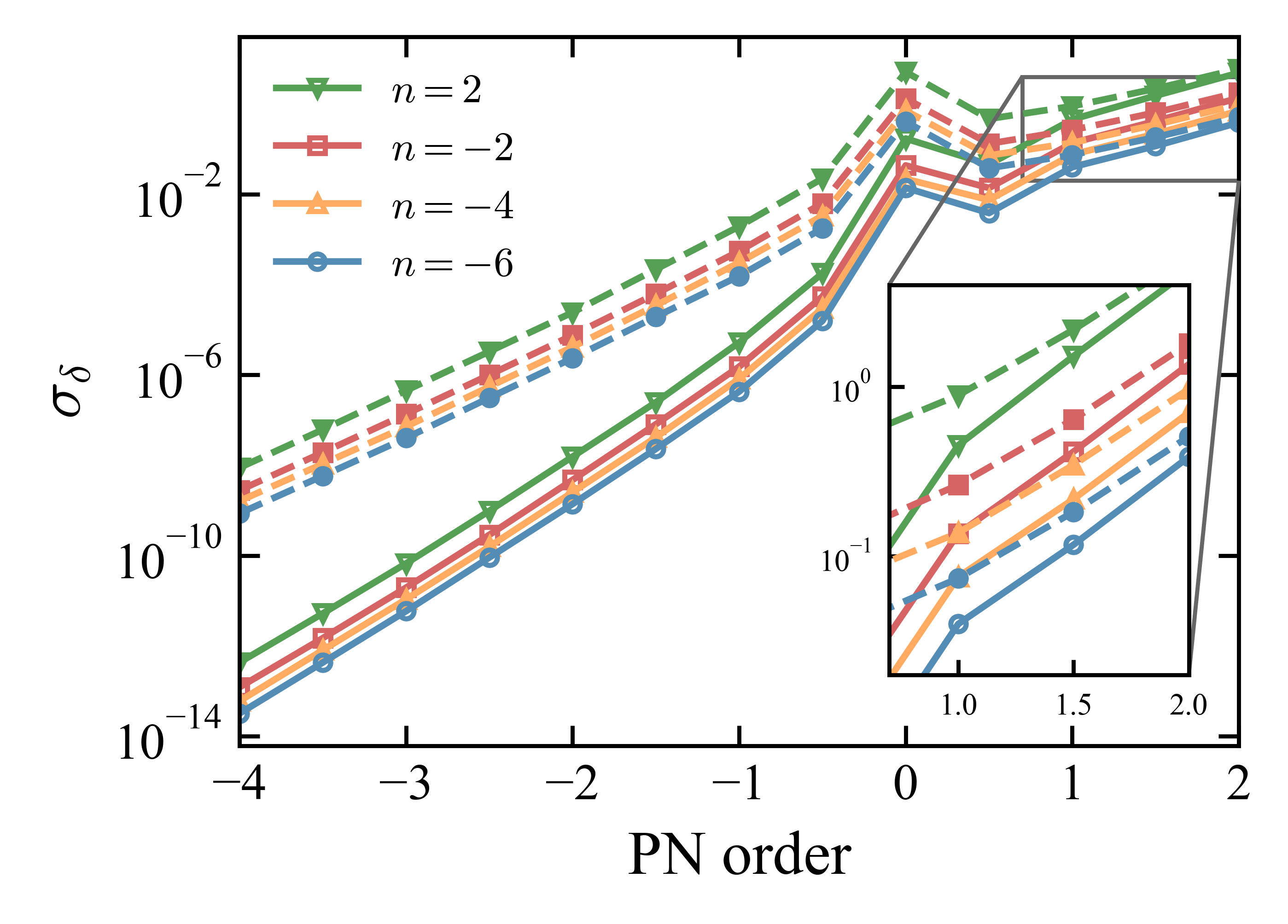}
        \caption{Constraints on the deformation-parameter uncertainty $\sigma_\delta$ as a function of PN order for different values of the index parameter $n$. \revised{Solid lines denote the space-based detector LISA, while dashed lines denote the ground-based detector LIGO.} Different colors correspond to different choices of $n$.}\label{fig:delta_PN_nppE}
    \end{minipage}
\end{figure}

\begin{figure}[ht]
    \begin{minipage}{\columnwidth}
        \centering
        \includegraphics[width=0.94\textwidth,
        trim=0 0 0 0,clip]{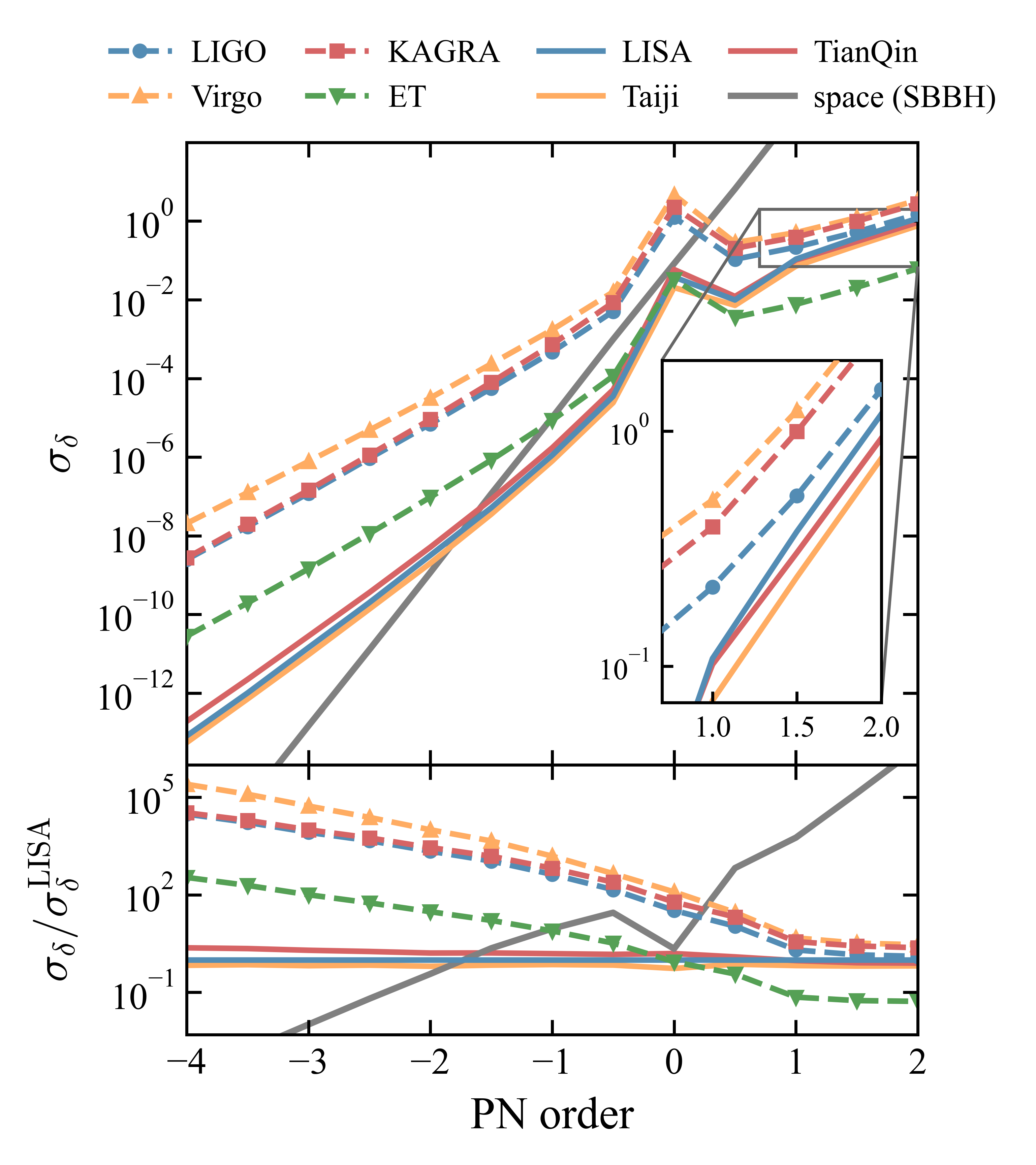}
        \caption{Constraints on the deformation-parameter uncertainty $\sigma_\delta$ as a function of PN order for different detectors. Solid lines denote space-based detectors, while dashed lines denote ground-based detectors. The gray solid line corresponds to the space--SBBH case; since the three space-based detectors yield nearly identical results in this scenario, we show only the LISA curve for clarity. The lower panel presents a relative comparison normalized to the LISA constraints.}\label{fig:delta_PN}
    \end{minipage}
\end{figure}

Under this assumption, the derivatives of the non-GR terms are simplified.
The logarithmic derivatives with respect to $\eta$ reduce to constants,
\begin{equation}
	\frac{\partial\ln\Xi}{\partial\ln\eta}
	=\frac{\partial\ln\Upsilon}{\partial\ln\eta}
	=\frac{n}{5},
\end{equation}
while derivatives with respect to $\mathcal M$ and $\chi_{s,a}$ vanish.
This allows a unified analysis of constraints associated with different PN-order corrections.
We consider PN orders in the range \mathrevised{n_{\rm PN}\in[-4,2]}, corresponding to $a\in[-8,4]$ and $b\in[-13,-1]$.
The index parameter is taken within $n\in[-6,2]$.
These choices encompass the dominant inspiral corrections predicted by most commonly studied modified gravity theories.
All remaining parameters are sampled according to Table~\ref{tab:parameters}.
Within this setup, Fisher-matrix analysis is performed for different detector configurations.
We systematically examine how the constraints depend on the PN order and the index parameter.
The results are shown in Figs.~\ref{fig:delta_PN_nppE} and~\ref{fig:delta_PN}.

\revised{To interpret the scaling behavior, we use the weak-correction, phase-dominated approximation adopted in Ref.}~\cite{ppE_TianQin}\revised{, in which the amplitude factor satisfies} \mathrevised{\mathcal R\simeq1}\revised{.}
The constraint on $\delta$ is then dominated by the $\Gamma_{\delta\delta}$ component,
\begin{equation}
	\Gamma_{\delta\delta}\simeq 
	\int_{f_{\min}}^{f_{\max}} 
	\frac{4|\tilde h|^2}{S_n(f)}
	\Big[(\Xi u^a)^2+(\Upsilon u^b)^2\Big]\mathrm{d} f .
\end{equation}
During the inspiral, one typically has $u<1$ and $a-b=5$.
The relative magnitude of the amplitude and phase contributions therefore satisfies
\begin{equation}
	\frac{u^{2a}}{u^{2b}}=u^{10}\ll 1 .
\end{equation}
The amplitude contribution is thus strongly suppressed.
For scaling arguments, we neglect this term and retain only the phase contribution.
We emphasize that this approximation is used solely for trend analysis, and all results presented in this work fully include the amplitude corrections.
Under this approximation, $\Gamma_{\delta\delta}$ reduces to
\begin{equation}\label{eq:Gamma_delta}
	\Gamma_{\delta\delta}\simeq 
	4 \mathcal A^2 \Upsilon^2
	\int_{f_{\min}}^{f_{\max}}
	\frac{f^{-7/3}}{S_n(f)}u^{2b}\mathrm{d} f .
\end{equation}

We first examine the dependence on the index parameter $n$.
Since $\sigma_\delta\propto\Gamma_{\delta\delta}^{-1/2}$, uncertainties at adjacent values of $n$ satisfy
\begin{equation}
	\frac{\sigma_\delta^{(n+1)}}{\sigma_\delta^{(n)}}\simeq \eta^{-1/5}.
\end{equation}
Because $\eta<1$, the uncertainty increases with increasing $n$.
This leads to a clear layered structure in Fig.~\ref{fig:delta_PN_nppE}.

The dominant parameter dependences in Eq.~(\ref{eq:Gamma_delta}) can be made explicit as
\begin{equation}
	\Gamma_{\delta\delta} \propto 
	\frac{\Upsilon^2  \mathcal M^{(5+2b)/3}}{D_L^2}
	\int_{f_{\min}}^{f_{\max}} 
	\frac{f^{(2b-7)/3}}{S_n(f)} \mathrm{d} f .
\end{equation}
This immediately implies a linear scaling of the uncertainty with distance,
\begin{equation}
	\sigma_\delta\propto D_L ,
\end{equation}
which reflects the standard distance dependence of the overall signal amplitude in the Fisher-matrix framework.
The dependence on PN order follows from $\sigma_\delta\propto\Gamma_{\delta\delta}^{-1/2}$.
For adjacent PN orders, the uncertainties approximately satisfy
\begin{equation}\label{eq:sigma_PN}
	\frac{\sigma_\delta^{(b+1)}}{\sigma_\delta^{(b)}}\simeq 
	\mathcal M^{-1/3}
	\left(
	\frac{\int \frac{f^{(2b-5)/3}}{S_n(f)} \mathrm{d}f}
	{\int \frac{f^{(2b-7)/3}}{S_n(f)} \mathrm{d}f}
	\right)^{-1/2}.
\end{equation}
This relation shows that changing the PN order affects the constraints not only through the explicit mass scaling, but also through the modified weighting of different frequency intervals in the Fisher integral. 
Such PN-dependent layering is clearly visible in Figs.~\ref{fig:delta_PN_nppE} and~\ref{fig:delta_PN}.

As shown in Fig.~\ref{fig:delta_PN}, most detectors exhibit a pronounced increase in uncertainty when the correction enters at 0PN order. 
This arises from the strong similarity between a 0PN correction and the leading GR term, which induces severe degeneracy with intrinsic binary parameters.
In contrast, this feature is absent for space-based observations of SBBHs (gray solid lines).
In this case, the signal spans only a narrow frequency interval and evolves slowly in the frequency domain.
Corrections at different PN orders therefore act as nearly constant rescalings rather than distinct shape modifications.
The dominant contribution to the Fisher matrix then arises from the overall SNR weighting, and the sensitivity to PN-order–dependent degeneracies is strongly suppressed. 
Consequently, no additional degeneracy is triggered at 0PN order.

Differences among detectors reflect their distinct sensitivity bands and observational characteristics. 
Among space-based detectors, Taiji provides the strongest overall constraints. 
Below 1PN, LISA outperforms TianQin due to its longer low-frequency integration.
Above 1PN, TianQin becomes increasingly competitive, reflecting its stronger response at higher frequencies.
For ground-based detectors, LIGO yields the strongest constraints among second-generation instruments.
The third-generation detector ET significantly outperforms them at all PN orders.
This improvement is driven by its lower starting frequency and enhanced sensitivity.

Overall, space-based detectors show a stronger dependence on PN order, while ground-based detectors exhibit a milder variation.
At higher PN orders, ground-based constraints approach those of space-based detectors, and ET can even surpass them.
This behavior highlights the complementarity between space-based and ground-based observations.

The analysis in this subsection focuses on the universal PN-order scaling of non-GR corrections.
In the next subsection, we examine how the constraints further depend on the intrinsic properties of the binary system.

\subsection{Dependence on Binary Properties}
\begin{figure*}[ht]
    \begin{minipage}{\textwidth}
        \centering
        \includegraphics[width=0.98\textwidth,
        trim=0 0 0 0,clip]{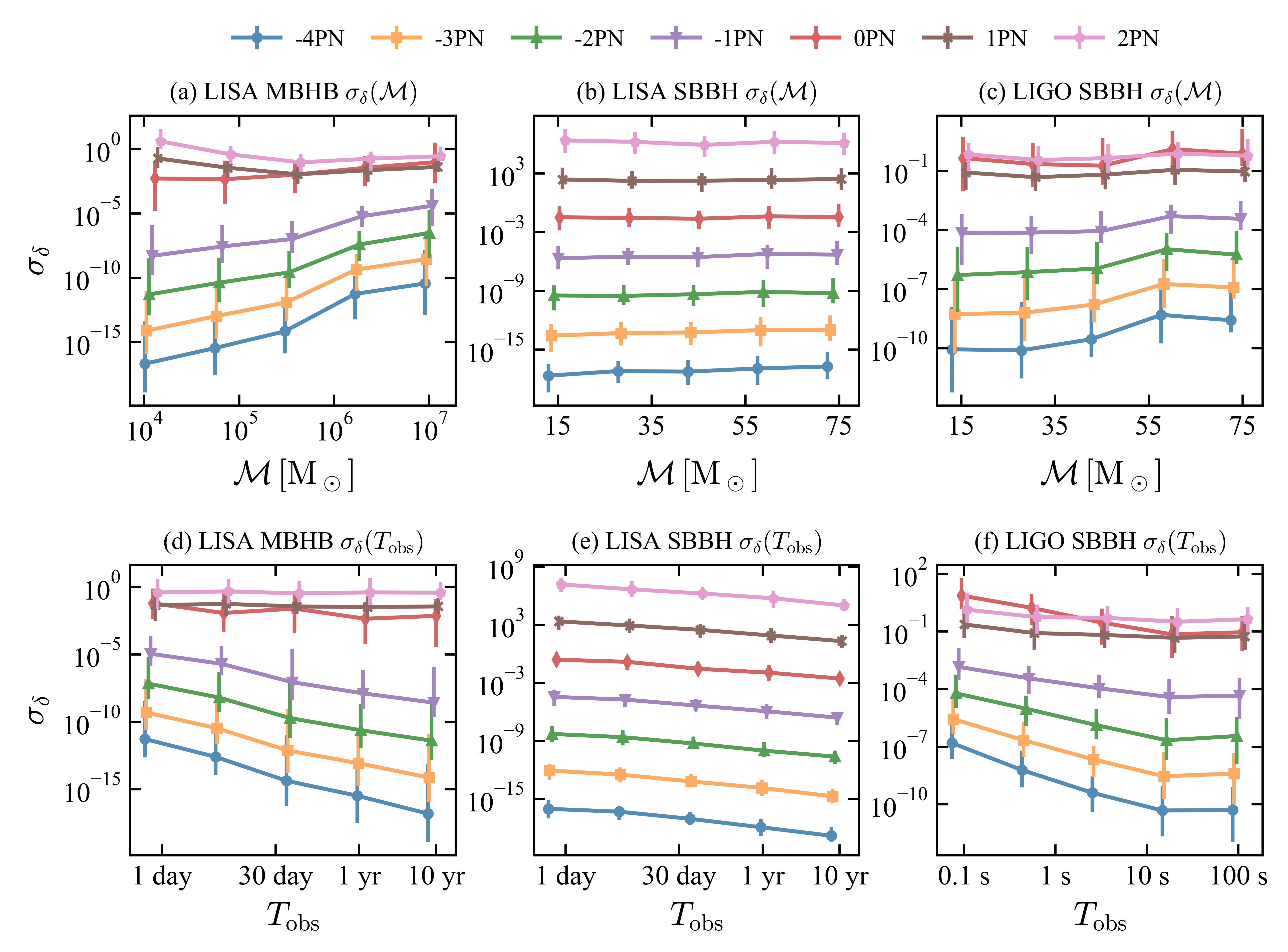}
        \caption{Dependence of the deformation-parameter uncertainty $\sigma_\delta$ on the chirp mass $\mathcal M$ and the observation time $T_{\rm obs}$. Panels (a)–(c) show $\sigma_\delta$ as a function of $\mathcal M$ for different detector and source configurations, where different colors and symbols correspond to non-GR corrections at different PN orders. Panels (d)–(f) display the corresponding dependence of $\sigma_\delta$ on the observation time $T_{\rm obs}$ under the same settings. Except for panels (b) and (c), which use linear scales on the horizontal axis, all other panels adopt logarithmic scales to more clearly illustrate the behavior over a wide dynamic range.}\label{fig:LISA_LIGO_M_T}
    \end{minipage}
\end{figure*}
After establishing the PN-order scaling of the constraints, we now examine their dependence on intrinsic BBH properties.
In a Fisher-matrix analysis, intrinsic parameters affect the waveform directly and also indirectly through the observable frequency range and its time evolution.
This modifies the frequency weighting in the Fisher integral and hence the final constraints.
In this subsection, we focus on the dependence of the deformation-parameter uncertainty $\sigma_\delta$ on the chirp mass $\mathcal M$ and the observation time $T_{\rm obs}$.
Following the previous subsection, we compute the constraints obtained by LISA and LIGO as functions of $\mathcal M$ and $T_{\rm obs}$.
The results are shown in Fig.~\ref{fig:LISA_LIGO_M_T}.

A robust feature is that corrections at different PN orders exhibit a layered structure, together with a pronounced transition at 0PN.
As implied by Eq.~(\ref{eq:sigma_PN}), the overall scaling between adjacent PN orders follows approximately $\mathcal M^{-1/3}$.
As $\mathcal M$ increases, the separation between different PN layers gradually decreases.
This behavior is clearly visible in Figs.~\ref{fig:LISA_LIGO_M_T}(a) and~(c).

For the space--MBHB and ground--SBBH cases, the frequency evolution is relatively rapid and the high-frequency cutoff is set by the ISCO.
The integration limits can be written as
\begin{equation}
\begin{aligned}
f_{\min}&=f(\tau=T_{\rm obs})
	\propto \mathcal M^{-5/8} T_{\rm obs}^{-3/8},\\
	f_{\max}&= f_{\rm ISCO}\propto M^{-1}.
\end{aligned}
\end{equation}

As the mass increases, both $f_{\min}$ and $f_{\max}$ shift to lower frequencies, with $f_{\max}$ decreasing more rapidly. 
The effective bandwidth therefore shrinks and the overall Fisher integral is reduced.
This leads to a degradation of the parameter constraints.
The detailed mass dependence is further modulated by the detector sensitivity curve and differs across PN orders.
In particular, for higher-PN corrections, the interplay between the shifting frequency band and the frequency-dependent sensitivity leads to a mass dependence that differs from that of lower-PN corrections.

In contrast, increasing the observation time $T_{\rm obs}$ leaves $f_{\max}$ unchanged in these scenarios while lowering $f_{\min}$.
\revised{Thus, for the space--MBHB and ground--SBBH cases, increasing} \mathrevised{T_{\rm obs}} \revised{corresponds to starting the inspiral observation earlier before the ISCO, i.e., at a lower frequency and a wider binary separation.}
The effective integration range is therefore extended toward lower frequencies.
As a result, the parameter uncertainty decreases monotonically with increasing $T_{\rm obs}$.
This trend is shown in Figs.~\ref{fig:LISA_LIGO_M_T}(d) and~(f).
For the ground--SBBH case, when $T_{\rm obs}$ becomes sufficiently large, the low-frequency extension may fall outside the sensitive band of LIGO.
In this regime, further increases in $T_{\rm obs}$ no longer enlarge the effective integration range.
The parameter uncertainty then approaches a constant value.

The space--SBBH case exhibits qualitatively different behavior.
Within the space-based detector band, the frequency evolution is extremely slow.
The observable frequency range can be approximated as
\begin{equation}
	f_{\min}= f_0,\qquad 
	f_{\max}= f_0+\Delta f,\qquad 
	\Delta f \ll f_0 .
\end{equation}
In this limit, Eq.~(\ref{eq:Gamma_delta}) reduces to
\begin{equation}
	\Gamma_{\delta\delta}\simeq
	4 \mathcal A^2 \Upsilon^2
	\frac{f_0^{-7/3}}{S_n(f_0)}
	(\pi\mathcal M f_0)^{2b/3} 
	\Delta f .
\end{equation}
Using the GW frequency evolution rate~\cite{Fisher_analytical},
\begin{equation}
	\dot f=\frac{96}{5}\pi^{8/3}\mathcal M^{5/3}f^{11/3},
\end{equation}
the frequency drift over $T_{\rm obs}$ can be approximated as $\Delta f\simeq \dot f T_{\rm obs}$.
\revised{For the space--SBBH case, by contrast, increasing} \mathrevised{T_{\rm obs}} \revised{extends the coherent observation forward from the same initial frequency} \mathrevised{f_0}\revised{, rather than shifting the template to an earlier starting frequency.}
This leads to the scaling relation
\begin{equation}
\begin{aligned}
	\sigma_\delta &\propto 
	\frac{D_L}{\Upsilon}
	\sqrt{\frac{S_n(f_0)}{T_{\rm obs}}}
	\mathcal M^{-(5+b)/3}
	f_0^{-(b+2)/3}\\
	&\propto 
	\mathrevised{\mathcal M^{-2n_{\rm PN}/3} T_{\rm obs}^{-1/2}}.
\end{aligned}
\end{equation}
\revised{Here} \mathrevised{b=2n_{\rm PN}-5} \revised{is the phase-correction exponent defined in Eq.~(7), and} \mathrevised{n_{\rm PN}} \revised{is the PN order of the ppE correction.}
The uncertainty therefore follows a simple and stable power-law dependence on $\mathcal M$ and $T_{\rm obs}$.
The corresponding trends are directly visible in Figs.~\ref{fig:LISA_LIGO_M_T}(b) and~(e).

In summary, the constraint on $\delta$ depends on BBH properties through both the signal bandwidth and the frequency weighting of the Fisher integral.
The chirp mass $\mathcal M$ reshapes the observable frequency distribution, while \revised{the observation time} \mathrevised{T_{\rm obs}} \revised{improves constraints either by extending the low-frequency coverage for rapidly evolving inspirals or by increasing the coherent integration time for slowly evolving space--SBBH signals.}
The relative importance of these effects varies across detector types and source classes.
This behavior provides a clear physical interpretation of the complementarity between space-based and ground-based detectors.
Together, they probe non-GR corrections at different PN orders in a complementary manner.

\section{Constraints on Specific non-GR Effects}\label{sec:Constraints_on_Specific}
In the preceding sections, we established a unified framework for constraining non-GR corrections in the ppE formalism across different PN orders, BBH parameters, and detector configurations.
We now apply this framework to a set of representative non-GR theories.
By mapping the leading corrections of each theory onto the ppE parameter space, we assess the ability of different detectors and source classes to constrain deviations from GR.

For each theory considered, the dominant inspiral correction is associated with a specific PN order and a corresponding set of ppE parameters.
Using the analytic waveform derivatives derived earlier, these parameters are constrained directly within the Fisher-matrix framework.
The ppE representations of the non-GR theories and their analytic derivatives are summarized in  \ref{app:ppE}.

\begin{figure*}[ht]
    \begin{minipage}{\textwidth}
        \centering
        \includegraphics[width=0.98\textwidth,
        trim=0 0 0 0,clip]{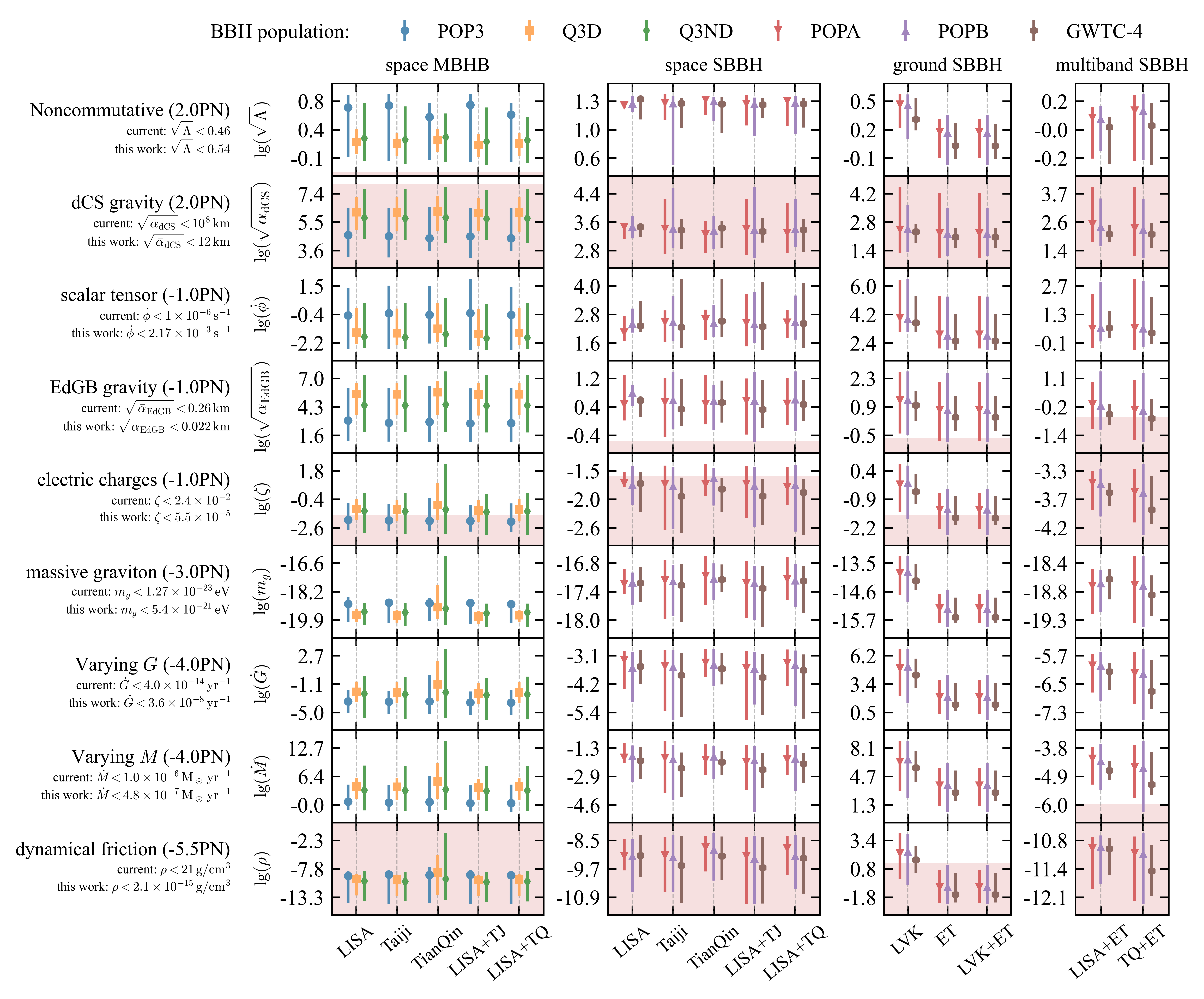}
        \caption{Comparison of constraints on non-GR parameters obtained by different detectors for various non-GR theories. The scatter points indicate the median constraints computed under the corresponding BBH population models, while the error bars represent the range between the maximum and minimum constraints across different population assumptions. The red shaded regions indicate constraints from existing observations. For theories without a red shaded region, our GW observations do not yet provide stronger constraints than those obtained from current observations. For each non-GR theory listed on the left, \textit{current} indicates the constraint level from current observations, whereas \textit{this work} shows the best constraint achieved in this study under the optimal detector and source configuration.}\label{fig:different_gravity}
    \end{minipage}
\end{figure*}

Our analysis covers MBHBs and SBBHs observed by space-based detectors, SBBHs observed by ground-based detectors, and multiband observations of SBBHs, as described in Sec.~\ref{subsec:detectors}. 
Astrophysical uncertainties are incorporated through several BBH population models, as discussed in Sec.~\ref{subsec:population}. 
The resulting constraints are summarized in Fig.~\ref{fig:different_gravity}, which serves as the main reference for this section.

In the following, we classify non-GR theories according to the PN order of their leading corrections.
We proceed from higher to lower PN orders and compare the relative performance of different detectors and source classes.

\paragraph{2PN non-GR corrections:}
Corrections entering at 2PN correspond to high-frequency modifications of the waveform phase.
Their detectability mainly relies on resolving fine structures at high frequencies rather than long-term phase accumulation.
As shown in Fig.~\ref{fig:different_gravity}, the differences among detector types are therefore relatively modest.

Noncommutative gravity introduces a noncommuting structure of spacetime coordinates in an attempt to alleviate ultraviolet divergences in quantum field theory at the level of gravity~\cite{Noncommutative_Gravity}. 
In this case, the strongest constraints are obtained from multiband--SBBH observations. 
This is because multiband measurements preserve high-frequency information while simultaneously introducing additional low-frequency constraints on the mass and spin parameters, thereby effectively reducing the correlation between high-PN corrections and intrinsic parameters in the Fisher matrix. 
Current constraints from the GW190814 event are already comparable to our results~\cite{current_noncommutative}. 
This indicates that present GW observations are already close to the achievable upper limit for constraining noncommutative gravity.

Dynamical Chern--Simons (dCS) gravity introduces a parity-violating quadratic curvature term into the action, leading to a 2PN phase modification in the waveform~\cite{dCS}. 
The tightest constraints arise from ground--SBBH observations, reflecting the advantage of ground-based detectors at high frequencies. 
Current bounds on dCS gravity mainly arise from satellite-based experiments~\cite{current_dCS}. 
In comparison, future GW observations are expected to tighten the upper bound on $\sqrt{\bar{\alpha}_{\rm dCS}}$ from the current $\mathcal{O}(10^8 {\rm km})$ down to the $\mathcal{O}(10 {\rm km})$ level, representing an improvement of \revised{many orders of magnitude}. 
This demonstrates that, even for positive-PN corrections, GWs can provide substantial gains over existing constraints in certain theoretical frameworks.

\paragraph{-1PN non-GR corrections:}
Scalar--tensor theories and Einstein--dilaton--Gauss--Bonnet (EdGB) gravity both introduce leading-order negative-PN corrections to the waveform. 
Charged BBH systems also generate corrections at the -1PN order, but the effective parameter space is usually strongly suppressed by charge-neutralization mechanisms in realistic astrophysical environments~\cite{electric_charges}. 
These effects dominate at low frequencies and accumulate over long inspiral durations.

In scalar--tensor theories, one scalar field is introduced through a nonminimal coupling to gravity~\cite{st}. 
For this class of theories, constraints from MBHBs are significantly stronger than those from SBBHs, with space--MBHB providing the tightest bounds. 
The most stringent existing constraints come from orbital decay measurements of the supermassive BBH system OJ287~\cite{current_st}, which remain stronger than current GW-based limits. 
This indicates that GW observations mainly serve as a complementary and independent probe.

In EdGB gravity, the scalar field nonminimally couples to a quadratic curvature term in the action~\cite{EdGB}. 
The resulting constraint pattern differs from that of scalar--tensor theories, with SBBHs yielding stronger bounds.
This difference illustrates that theories entering at the same PN order can exhibit distinct behaviors due to their different parameter dependences.

Constraints on BH electric charge follow a pattern similar to EdGB gravity.
Multiband--SBBH provides the tightest bounds, while space--MBHB observations remain competitive. 
Current GW limits on both EdGB gravity and BH charge mainly come from GW230529~\cite{current_EdGB_ec}, and future GW observations are expected to further improve these constraints by several orders of magnitude.

\paragraph{-3PN non-GR corrections:}
A nonzero graviton mass enters the GW phase as a generation modification, appearing at the -3PN order and being particularly sensitive to low-frequency signals~\cite{mg}. 
Our results show that space--MBHB observations provide the strongest constraints in this case. 
Although SBBH observations are not disadvantaged in terms of SNR, their relatively high starting frequencies prevent efficient accumulation of low-frequency propagation effects, fundamentally limiting their sensitivity to the graviton mass.

The most stringent existing bounds on the graviton mass are obtained from GWTC--3~\cite{GWTC3_test}.
These limits are already close to the sensitivity achievable with GW observations, leaving limited room for further improvement.

\paragraph{-4PN non-GR corrections:}
Time-varying $G$ and Time-varying $M$ theories consider the slowly changing gravitational constant and BBH mass over time~\cite{test_GR1,gdot,mdot}.
They introduce even lower-PN corrections to the waveform, corresponding to slow, monotonic shifts in the phase evolution.

Our results indicate that multiband--SBBH provides the strongest constraints for both theories, while space--SBBH observations are also competitive, reflecting the advantage of low-frequency observations in constraining long-term secular effects. 
Current bounds on $\dot{G}$ from the NASA MESSENGER mission~\cite{current_gdot} remain stronger than GW-based limits. 
In contrast, constraints on $\dot{M}$ are largely theoretical, and future GW observations are expected to provide meaningful bounds.

\paragraph{-5.5PN non-GR corrections:}
Dynamical friction represents the lowest-PN correction considered in this work, with its effect being almost entirely dominated by long-term dissipative accumulation at low frequencies~\cite{dynamical_friction}. 
Space-based detectors outperform ground-based detectors by more than ten orders of magnitude in this case. 
This dramatic difference arises from the long observation times of space-based detectors.
Such observations allow extremely weak dissipative effects to build up into measurable phase shifts.

These results clearly demonstrate that, for ultra-low-PN environmental or dissipative effects, space-based GW observations possess an intrinsic and irreplaceable advantage. 
This highlights their unique potential for probing the environments of \revised{compact binary systems}. 
\revised{Current constraints on compact binary environments} mainly come from GW170817~\cite{current_df}. In addition to space-based detectors, future observations with ET are expected to further improve the constraints by nearly four orders of magnitude.

Overall, the constraining power of GW observations on non-GR corrections is largely determined by the PN order of the correction. 
Positive-PN corrections mainly rely on phase resolution at high frequencies, leading to relatively small differences among detector configurations. 
In contrast, negative-PN corrections accumulate coherently during the low-frequency inspiral, making low-frequency coverage and observation time the dominant factors. 

It is worth noting that, even at the same PN order, different theories can display distinct constraint trends due to differences in how their corrections correlate with intrinsic binary parameters. 
GW observations therefore provide unique access to new parameter regimes, particularly for low-PN and environment-related effects.

\section{Conclusions}\label{sec:Conclusions}
In this work, we derived fully analytic waveform derivatives within the Fisher-matrix formalism based on the ppE framework.
Using the frequency-domain \texttt{TaylorF2} waveform, we constructed a unified parameterized model and developed an analytic Fisher-matrix approach.
This framework was applied to a broad class of non-GR theories to compute and compare the constraining power of GW observations.
The analytic approach enables efficient evaluation of waveform responses to parameter variations over the full parameter space.
It also allows a consistent treatment of different BBH population models within the same framework.
Based on this setup, we performed a unified analysis of multiple modified gravity theories and environmental effects under various detector configurations, source classes, and multiband observation scenarios.

From a methodological perspective, the analytic waveform derivatives exhibit clear advantages in numerical stability and computational efficiency.
Compared with the finite-difference methods, the analytic approach substantially reduces the computational cost of individual Fisher-matrix evaluations.
It also avoids systematic uncertainties associated with step-size choices.
These advantages become increasingly important in applications that require repeated Fisher-matrix calculations, such as large-scale parameter scans and more complex data-analysis pipelines.
The analytic framework therefore provides an efficient and consistent tool for assessing the capability of future GW observations to test non-GR theories.

From a physical standpoint, the constraining power on non-GR effects is primarily determined by the PN order at which the correction enters the waveform. 
Positive-PN corrections mainly rely on high-frequency waveform information and show relatively weak sensitivity to detector configuration. 
In contrast, negative-PN corrections accumulate coherently during the low-frequency inspiral, making low-frequency coverage and long observation times the key factors for improving constraints. 
In this regime, space-based detectors and multiband observations exhibit clear advantages, and in some cases are capable of achieving order-of-magnitude improvements over existing observational bounds.

The non-GR effects considered in this work mainly modify the GW generation mechanism.
The analytic framework developed here can be extended to propagation-related effects, such as dispersive propagation or higher-PN phase corrections.
Moreover, combining analytic waveform derivatives with parameter-estimation techniques such as Markov chain Monte Carlo methods represents a promising direction.
Such extensions would enable more comprehensive statistical inference of non-GR parameters while retaining computational efficiency.
We plan to pursue these extensions in future work.

\begin{acknowledgements}
This work was supported by the National Key Research and Development Program of China (Grant No. 2023YFC2206702), the National Natural Science Foundation of China (Grant Nos. 125B2102, 12575072, and 12547101), the Fundamental Research Funds for the Central Universities Project (Grant No. 2024IAIS-ZD009), and the Natural Science Foundation of Chongqing (Grant No. CSTB2023NSCQ-MSX0103).
\end{acknowledgements}

\appendix
\section{PN phase coefficients and their derivatives}\label{app:PN_phase}
In this appendix, we present the PN phase expansion adopted in our analysis, together with the explicit analytic derivatives of the phase with respect to the mass and spin parameters. 
These results provide the technical basis for the analytic Fisher-matrix calculations in the main text and are collected here for convenience. 
Under the SPA, the GR phase of the BBH inspiral can be written as the PN power-series expansion~\cite{PN_coefficients}
\begin{equation}
	\Psi_{\rm GR}(f)
	= 2\pi f t_c - \phi_c - \frac{\pi}{4}
	+ \frac{3}{128\eta} \sum_{p=0}^{4} \varphi_p v^{p-5} .
\end{equation}
In the main text, we truncate the PN expansion at the 2PN order. The corresponding PN coefficients are
\begin{equation}
	\varphi_0=1,\qquad \varphi_1=0,\qquad
	\varphi_2=\frac{3715}{756}+\frac{55}{9}\eta,
\end{equation}
\begin{equation}
	\varphi_3=-16\pi+\frac{113}{3}\chi_a\xi
	+\left(\frac{113}{3}-\frac{76}{3}\eta\right)\chi_s,
\end{equation}
\begin{equation}
\begin{aligned}
	\varphi_4=\frac{15293365}{508032}
	+\frac{27145}{504}\eta+\frac{3085}{72}\eta^2 \\
	+\left(-\frac{405}{8}+200\eta\right)\chi_a^2
	-\frac{405}{4}\chi_s\chi_a\xi \\
	+\left(-\frac{405}{8}+\frac{5}{2}\eta\right)\chi_s^2.
\end{aligned}
\end{equation}
Here $\xi = (m_1-m_2)/M = \sqrt{1-4\eta}$ is the mass asymmetry parameter $(m_1\ge m_2)$.

The Fisher-matrix evaluation requires the partial derivatives of the phase with respect to the model parameters. 
We first consider the mass parameters. 
Since the masses enter the phase both through the overall frequency scaling and through the PN coefficients, the derivatives contain multiple contributions. 
For the chirp mass $\mathcal M$ and the symmetric mass ratio $\eta$, we obtain
\begin{equation}
	\frac{\partial\Psi_{\rm GR}}{\partial\ln\mathcal M}
	=\frac{1}{128 \eta}
	\sum_{p=0}^{4}(p-5) \varphi_p v^{ p-5},
\end{equation}
\begin{equation}
	\frac{\partial\Psi_{\rm GR}}{\partial\ln\eta}
	=\frac{3}{128 \eta}
	\sum_{p=0}^{4}
	\left[
	\frac{\partial\varphi_p}{\partial\ln\eta}
	-\frac{p}{5}\varphi_p
	\right]v^{ p-5}.
\end{equation}
It can be seen that $\varphi_p$ depends only on $\eta$, but not on $\mathcal{M}$. 
Considering that $\partial\xi/\partial\ln\eta=-2\eta/\xi$, we obtain
\begin{equation}
	\frac{\partial\varphi_0}{\partial\ln\eta}=
	\frac{\partial\varphi_1}{\partial\ln\eta}=0,\qquad
	\frac{\partial\varphi_2}{\partial\ln\eta}=\eta \frac{55}{9},
\end{equation}
\begin{equation}
	\frac{\partial\varphi_3}{\partial\ln\eta}
	=-\frac{226}{3}\frac{\eta\chi_a}{\xi}-\frac{76}{3}\eta\chi_s,
\end{equation}
\begin{equation}
	\frac{\partial\varphi_4}{\partial\ln\eta}
	=\eta\left(\frac{27145}{504}+\frac{3085}{36}\eta
	+200 \chi_a^2+\frac{5}{2}\chi_s^2\right)
	+\frac{405}{2}\frac{\eta\chi_s\chi_a}{\xi}.
\end{equation}

We next consider the spin dependence. Spin effects mainly enter the higher-PN terms of the phase through the symmetric and antisymmetric spin combinations $\chi_s$ and $\chi_a$. The phase derivatives with respect to the spin parameters are given by
\begin{equation}
	\frac{\partial\Psi_{\rm GR}}{\partial\chi_{s,a}}
	=\frac{3}{128 \eta}
	\sum_{p=0}^{4}
	\frac{\partial\varphi_p}{\partial\chi_{s,a}} 
	v^{ p-5},
\end{equation}
with
\begin{equation}
	\frac{\partial\varphi_1}{\partial\chi_{s}}
	=\frac{\partial\varphi_2}{\partial\chi_{s}}=\frac{\partial\varphi_1}{\partial\chi_{a}}
	=\frac{\partial\varphi_2}{\partial\chi_{a}}=0,
\end{equation}
\begin{equation}
	\frac{\partial\varphi_3}{\partial\chi_s}=\frac{113}{3}-\frac{76}{3}\eta,\qquad
	\frac{\partial\varphi_3}{\partial\chi_a}=\frac{113}{3}\xi,
\end{equation}
\begin{equation}
	\frac{\partial\varphi_4}{\partial\chi_s}
	=-\frac{405}{4}\chi_a\xi
	+2\left(-\frac{405}{8}+\frac{5}{2}\eta\right)\chi_s,
\end{equation}
\begin{equation}
	\frac{\partial\varphi_4}{\partial\chi_a}
	=2\left(-\frac{405}{8}+200\eta\right)\chi_a
	-\frac{405}{4}\chi_s\xi.
\end{equation}

We emphasize that the above derivatives involve only the GR phase $\Psi_{\rm GR}$. In the Fisher-matrix analysis in the main text, derivatives of the non-GR correction terms are treated separately within the ppE framework. Their explicit forms depend on the specific beyond-GR scenario under consideration and are listed in  \ref{app:ppE}. Combining the results in this appendix with those in  \ref{app:ppE} yields a fully analytic set of waveform derivatives with respect to all parameters.

\section{Analytic derivatives of non-GR corrections}\label{app:ppE}
This appendix systematically lists all inputs required for the analytic evaluation of waveform derivatives in the ppE framework for the non-GR theories considered in this work. 
We emphasize that our goal is not to re-derive or review the complete theoretical structure of each modified gravity theory, but rather to derive the analytic waveform derivatives within the ppE framework based on existing results in the literature.

The ppE amplitude and phase corrections adopted here are primarily taken from Refs.~\cite{ppE_both_A_Psi,ppE_only_Psi}. 
The case of electrically charged BHs follows Ref.~\cite{electric_charges}, while the dynamical-friction effects induced by environmental matter are based on Ref.~\cite{dynamical_friction}. 
For detailed theoretical backgrounds and derivations, we refer the reader to these references. 
In this appendix, we present only the mathematical expressions that are essential for our analysis.

As discussed in Sec.~\ref{sec:Fisher_matrix}, for a given non-GR theory it is sufficient to specify the quantities $\delta$, $\Xi$, and $\Upsilon$, together with the derivatives
\begin{equation}
\frac{\partial\ln\Xi}{\partial\ln\mathcal M},
\frac{\partial\ln\Upsilon}{\partial\ln\mathcal M},
\frac{\partial\ln\Xi}{\partial\ln\eta},
\frac{\partial\ln\Upsilon}{\partial\ln\eta},
\frac{\partial\ln\Xi}{\partial \chi_{s,a}},
\frac{\partial\ln\Upsilon}{\partial \chi_{s,a}},
\end{equation}
to systematically construct the analytic derivatives of the waveform with respect to all parameters within the unified ppE--Fisher framework. 
If a given correction does not depend on a specific parameter, the corresponding derivative vanishes identically. 
In the following, we present the specific non-GR theories adopted in this work and their complete sets of derivatives, ordered by the PN order at which the leading correction enters, from higher to lower PN orders.

\subsection{Noncommutative gravity}
Noncommutative gravity modifies the conservative dynamics of BBH systems and introduces a correction at the 2PN order during the inspiral. In the ppE framework adopted in this work, this theory is parameterized by
\begin{equation}
\delta=\Lambda^2,\qquad
a=4,\qquad
b=-1,
\end{equation}
where $\sqrt{\Lambda}$ characterizes the scale of quantum spacetime.

In this theory, both the amplitude correction function $\Xi$ and the phase correction function $\Upsilon$ depend only on the symmetric mass ratio $\eta$, with explicit forms given by
\begin{equation}
\Xi_{\rm NC}
=-\frac{3}{8}\eta^{-4/5}(2\eta-1),
\end{equation}
\begin{equation}
\Upsilon_{\rm NC}
=-\frac{75}{256}\eta^{-4/5}(2\eta-1).
\end{equation}
From these expressions, the logarithmic derivatives with respect to $\eta$ follow directly as
\begin{equation}
\frac{\partial\ln\Xi_{\rm NC}}{\partial\ln\eta}
=\frac{\partial\ln\Upsilon_{\rm NC}}{\partial\ln\eta}
=-\frac{4}{5}+\frac{2\eta}{2\eta-1}.
\end{equation}
All other logarithmic derivatives with respect to the waveform parameters vanish.

\subsection{dCS gravity}
Dynamical Chern--Simons gravity also introduces a 2PN correction during the inspiral phase, but in this case the effect depends sensitively on the spin structure of the BBH system. In the ppE framework adopted in this work, we take
\begin{equation}
\delta=\bar{\alpha}_{\rm dCS}^{2},\qquad
a=4,\qquad b=-1,
\end{equation}
where $\bar{\alpha}_{\rm dCS}$ denotes the coupling parameter. Constraints are conventionally quoted in terms of $\sqrt{\bar{\alpha}_{\rm dCS}}$.

For notational convenience, we define
\begin{equation}
K_\alpha=\frac{57713}{344064},\qquad
K_\beta=\frac{481525}{3670016},
\end{equation}
together with the spin-dependent combinations
\begin{equation}
S_\alpha=-2 \xi \chi_a\chi_s
+\left(1-\frac{14976 \eta}{57713}\right)\chi_a^2
+\left(1-\frac{215876 \eta}{57713}\right)\chi_s^2,
\end{equation}
\begin{equation}
S_\beta=-2 \xi \chi_a\chi_s
+\left(1-\frac{4992 \eta}{19261}\right)\chi_a^2
+\left(1-\frac{72052 \eta}{19261}\right)\chi_s^2.
\end{equation}
Using $M=\mathcal M\,\eta^{-3/5}$, the corresponding amplitude and phase correction functions can then be written as
\begin{equation}
\Xi_{\rm dCS}
=\frac{16\pi}{\mathcal{M}^4} \eta^{-2/5} K_\alpha S_\alpha,
\qquad
\Upsilon_{\rm dCS}
=\frac{16\pi}{\mathcal{M}^4} \eta^{-2/5} K_\beta S_\beta.
\end{equation}

We next compute the derivatives of the waveform with respect to the intrinsic parameters. For the chirp mass, one finds
\begin{equation}
\frac{\partial\ln\Xi_{\rm dCS}}{\partial\ln\mathcal{M}}
=\frac{\partial\ln\Upsilon_{\rm dCS}}{\partial\ln\mathcal{M}}
=-4.
\end{equation}
The derivatives with respect to the symmetric mass ratio are given by
\begin{equation}
\frac{\partial\ln\Xi_{\rm dCS}}{\partial\ln\eta}
=-\frac{2}{5}
+\frac{\eta}{S_\alpha}\left(
\frac{4}{\xi}\chi_a\chi_s
-\frac{14976}{57713} \chi_a^2
-\frac{215876}{57713} \chi_s^2
\right),
\end{equation}
\begin{equation}
\frac{\partial\ln\Upsilon_{\rm dCS}}{\partial\ln\eta}
=-\frac{2}{5}
+\frac{\eta}{S_\beta}\left(
\frac{4}{\xi}\chi_a\chi_s
-\frac{4992}{19261} \chi_a^2
-\frac{72052}{19261} \chi_s^2
\right).
\end{equation}
Finally, the derivatives with respect to the spin parameters read
\begin{equation}
\frac{\partial\ln\Xi_{\rm dCS}}{\partial\chi_s}
=\frac{-2\xi \chi_a
+2\left(1-\frac{215876 \eta}{57713}\right)\chi_s}{S_\alpha},
\end{equation}
\begin{equation}
\frac{\partial\ln\Xi_{\rm dCS}}{\partial\chi_a}
=\frac{-2\xi \chi_s
+2\left(1-\frac{14976 \eta}{57713}\right)\chi_a}{S_\alpha},
\end{equation}
\begin{equation}
\frac{\partial\ln\Upsilon_{\rm dCS}}{\partial\chi_s}
=\frac{-2\xi \chi_a
+2\left(1-\frac{72052 \eta}{19261}\right)\chi_s}{S_\beta},
\end{equation}
\begin{equation}
\frac{\partial\ln\Upsilon_{\rm dCS}}{\partial\chi_a}
=\frac{-2\xi \chi_s
+2\left(1-\frac{4992 \eta}{19261}\right)\chi_a}{S_\beta}.
\end{equation}

\subsection{EdGB gravity}
EdGB gravity introduces a leading $-1$PN correction during the inspiral phase through dipole radiation. In the ppE framework adopted in this work, we take
\begin{equation}
\delta=\bar{\alpha}_{\rm EdGB}^{ 2},\qquad
a=-2,\qquad b=-7,
\end{equation}
where $\bar{\alpha}_{\rm EdGB}$ is the coupling parameter, defined in close analogy to $\bar{\alpha}_{\rm dCS}$.

We introduce the dimensionless EdGB black-hole scalar charge
\begin{equation}
\tilde s(\chi_i)=
\frac{2\left(\sqrt{1-\chi_i^2}-1+\chi_i^2\right)}{\chi_i^2},
\end{equation}
and define the combination
\begin{equation}
B=(1+\xi)^2 \tilde s(\chi_2)-(1-\xi)^2 \tilde s(\chi_1).
\end{equation}
The amplitude and phase correction functions can then be written as
\begin{equation}
\Xi_{\rm EdGB}
=-\frac{5\pi}{192} 
\frac{B^2}{\mathcal{M}^4 \eta^{6/5}},\qquad
\Upsilon_{\rm EdGB}
=-\frac{5\pi}{7168} 
\frac{B^2}{\mathcal{M}^4 \eta^{6/5}}.
\end{equation}

The logarithmic derivatives with respect to the chirp mass are identical to those in dCS gravity,
\begin{equation}
\frac{\partial\ln\Xi_{\rm EdGB}}{\partial\ln\mathcal{M}}
=\frac{\partial\ln\Upsilon_{\rm EdGB}}{\partial\ln\mathcal{M}}
=-4.
\end{equation}
For the symmetric mass ratio, one finds
\begin{equation}
\frac{\partial\ln\Xi_{\rm EdGB}}{\partial\ln\eta}
=\frac{\partial\ln\Upsilon_{\rm EdGB}}{\partial\ln\eta}
=-\frac{6}{5}
-\frac{4\eta}{B \xi} 
\frac{\partial B}{\partial\xi}.
\end{equation}
Here,
\begin{equation}
\frac{\partial B}{\partial\xi}
=2(1+\xi)\tilde s(\chi_2)
+2(1-\xi)\tilde s(\chi_1).
\end{equation}

The derivatives with respect to the spin parameters satisfy
\begin{equation}
\frac{\partial\ln\Xi_{\rm EdGB}}{\partial\chi_s}
=\frac{\partial\ln\Upsilon_{\rm EdGB}}{\partial\chi_s}
=\frac{2}{B}\frac{\partial B}{\partial\chi_s},
\end{equation}
\begin{equation}
\frac{\partial\ln\Xi_{\rm EdGB}}{\partial\chi_a}
=\frac{\partial\ln\Upsilon_{\rm EdGB}}{\partial\chi_a}
=\frac{2}{B}\frac{\partial B}{\partial\chi_a},
\end{equation}
with
\begin{equation}
\frac{\partial B}{\partial\chi_s}
=(1+\xi)^2 \tilde s_2'-(1-\xi)^2 \tilde s_1',
\end{equation}
\begin{equation}
\frac{\partial B}{\partial\chi_a}
=-(1+\xi)^2 \tilde s_2'-(1-\xi)^2 \tilde s_1'.
\end{equation}
The derivative of the scalar charge with respect to the spin reads
\begin{equation}
\tilde s_i'=\frac{{\rm d}\tilde s}{{\rm d}\chi_i}
=-\frac{2}{\chi_i\sqrt{1-\chi_i^2}}
-\frac{4\left(\sqrt{1-\chi_i^2}-1\right)}{\chi_i^3}.
\end{equation}

\subsection{Scalar--tensor gravity}
Scalar--tensor gravity also introduces a leading $-1$PN correction during the inspiral phase. In the ppE framework adopted in this work, we take
\begin{equation}
\delta=\dot{\phi}^{ 2},\qquad
a=-2,\qquad b=-7,
\end{equation}
where $\dot{\phi}$ denotes the growth rate of the scalar field.

We introduce the dimensionless scalar charge
\begin{equation}
s(\chi_i)=\frac{1+\sqrt{1-\chi_i^2}}{2},
\end{equation}
and define the combination
\begin{equation}
A=\bigl[s(\chi_1)-s(\chi_2)\bigr]
+\xi \bigl[s(\chi_1)+s(\chi_2)\bigr].
\end{equation}
The corresponding amplitude and phase correction functions are then given by
\begin{equation}
\Xi_{\rm ST}
=-\frac{5}{192} \mathcal{M}^{2} \eta^{-4/5} A^{2},
\qquad
\Upsilon_{\rm ST}
=-\frac{5}{7168} \mathcal{M}^{2} \eta^{-4/5} A^{2}.
\end{equation}

The logarithmic derivatives with respect to the chirp mass read
\begin{equation}
\frac{\partial\ln\Xi_{\rm ST}}{\partial\ln\mathcal{M}}
=\frac{\partial\ln\Upsilon_{\rm ST}}{\partial\ln\mathcal{M}}
=2.
\end{equation}
For the symmetric mass ratio, one finds
\begin{equation}
\frac{\partial\ln\Xi_{\rm ST}}{\partial\ln\eta}
=\frac{\partial\ln\Upsilon_{\rm ST}}{\partial\ln\eta}
=-\frac{4}{5}
+\frac{2}{A} \frac{\partial A}{\partial\ln\eta},
\end{equation}
where
\begin{equation}
\frac{\partial A}{\partial\ln\eta}
=-\frac{2\eta}{\xi}
\bigl[s(\chi_1)+s(\chi_2)\bigr].
\end{equation}

The derivatives with respect to the spin parameters satisfy
\begin{equation}
\frac{\partial\ln\Xi_{\rm ST}}{\partial\chi_s}
=\frac{\partial\ln\Upsilon_{\rm ST}}{\partial\chi_s}
=\frac{2}{A} \frac{\partial A}{\partial\chi_s},
\end{equation}
\begin{equation}
\frac{\partial\ln\Xi_{\rm ST}}{\partial\chi_a}
=\frac{\partial\ln\Upsilon_{\rm ST}}{\partial\chi_a}
=\frac{2}{A} \frac{\partial A}{\partial\chi_a}.
\end{equation}
Here,
\begin{equation}
\frac{\partial A}{\partial\chi_s}
=(s_1'-s_2')+\xi (s_1'+s_2'),
\end{equation}
\begin{equation}
\frac{\partial A}{\partial\chi_a}
=(s_1'+s_2')+\xi (s_1'-s_2'),
\end{equation}
with
\begin{equation}
s_i'
=\frac{{\rm d}s}{{\rm d}\chi_i}
=-\frac{\chi_i}{2\sqrt{1-\chi_i^2}}.
\end{equation}

\subsection{Black-hole electric charges}
Electric charges carried by BHs can introduce additional radiation channels during the inspiral, leading to $-1$PN non-GR correction. In the ppE framework adopted in this work, this effect is parameterized by
\begin{equation}
\delta=\zeta^2,\qquad
b=-7,
\end{equation}
where $\zeta$ denotes the difference between the charges of the BBH. In this model, the leading-order correction is purely in the phase sector, i.e. $\alpha_{\zeta}=0$.

Accordingly, the amplitude correction function vanishes,
\begin{equation}
\Xi_{\rm charge}=0,
\end{equation}
while the phase correction function depends only on the symmetric mass ratio $\eta$ and can be written as
\begin{equation}
\Upsilon_{\rm charge}
=-\frac{5}{3584} \eta^{2/5} \kappa_{\rm ec}^{-1/3}.
\end{equation}
Here $\kappa_{\rm ec}=1-\lambda_1 \lambda_2\approx 1$ since we consider $\lambda_i=q_i/m_i$ as small perturbations.

From the above expression, the only nonvanishing logarithmic derivative is with respect to $\eta$,
\begin{equation}
\frac{\partial\ln\Upsilon_{\rm charge}}{\partial\ln\eta}=\frac{2}{5},
\end{equation}
whereas all other logarithmic derivatives with respect to the waveform parameters vanish.

\subsection{Dynamical graviton mass}
Dynamical graviton mass theory enters the waveform as a generation modification, leading to a -3PN correction in the inspiral phase. In the ppE framework adopted in this work, the effect of a dynamical graviton mass is parameterized as
\begin{equation}
\delta=(m_g/\hbar)^2,\qquad
b=-11,
\end{equation}
where $m_g$ denotes the dynamical graviton mass. At leading order, this modification enters only through the phase sector of the waveform.

Accordingly, the amplitude correction function vanishes,
\begin{equation}
\Xi_{\rm mg}=0,
\end{equation}
while the phase correction function is independent of the intrinsic binary parameters and can be written as
\begin{equation}
\Upsilon_{m_g}=\frac{25}{19712} \frac{\mathcal M^2}{F(e)},
\end{equation}
where $F(e)$ is a function of the eccentricity, taken to be 1 for our work.

As a result, all logarithmic derivatives of the correction functions with respect to the waveform parameters vanish, except for 
\begin{equation}
	\frac{\partial\ln\Upsilon_{m_g}}{\partial\ln\mathcal M}=2.
\end{equation}

\subsection{Time-varying $G$ theory}
In theories with a time-varying gravitational constant, the secular evolution of $G$ modifies the conservative dynamics and radiation reaction of compact binaries, leading to a -4PN correction in the inspiral phase. In the ppE framework adopted in this work, the leading-order effect of a varying $G$ is parameterized as
\begin{equation}
\delta={\dot G},\qquad
a=-8,\qquad
b=-13,
\end{equation}
where $\dot G$ characterizes the fractional rate of change of the gravitational constant.
The full ppE parameters are given by
\begin{equation}
\begin{aligned}
	\alpha_{\dot G}=
\frac{5}{512} \eta_0^{3/5} \dot G_{C,0}
\Big[
-7m_0+(s_{1,0}+s_{2,0}-\delta_{\dot G})m_0\\
+13(m_{1,0}s_{1,0}+m_{2,0}s_{2,0})
\Big],
\end{aligned}
\end{equation}
\begin{equation}
\begin{aligned}
	\beta_{\dot G}=
-\frac{25}{851968} \eta_0^{3/5} \dot G_{C,0}
\Big[
11m_0+3(s_{1,0}+s_{2,0}-\delta_{\dot G})m_0\\
-41(m_{1,0}s_{1,0}+m_{2,0}s_{2,0})
\Big].
\end{aligned}
\end{equation}
Here we consider $G_D=G_C=0$, which corresponds to $\delta_{\dot G}=0$, and set $s_i=0.5$ for each BH.
The subscript $0$ denotes that the quantity is measured at the time $t=t_0$. Thus the expressions of the ppE parameters can be written as
\begin{equation}
	\Xi_{\dot G}=\frac{5}{512} \mathcal M F_\alpha,
\end{equation}
\begin{equation}
	\Upsilon_{\dot G}=-\frac{25}{851968} \mathcal M F_\beta,
\end{equation}
where $F_\alpha$ and $F_\beta$ are the corresponding expressions. Under our settings, they can be simplified into constants
\begin{equation}
	F_\alpha=\frac{1}{2},\qquad F_\beta=-\frac{13}{2}.
\end{equation}

The logarithmic derivatives with respect to the chirp mass read
\begin{equation}
\frac{\partial\ln\Xi_{\dot G}}{\partial\ln\mathcal M}
=\frac{\partial\ln\Upsilon_{\dot G}}{\partial\ln\mathcal M}=1,
\end{equation}
the derivatives of the remaining items are 0.

\subsection{Time-varying $M$ theory}
A secular variation of the component masses modifies the inspiral dynamics by inducing an additional phase accumulation in the gravitational waveform. At leading order, this effect enters the phase at the -4PN order. In the ppE framework adopted in this work, the time-varying mass effect is parameterized as
\begin{equation}
\delta={\dot M},\qquad
b=-13,
\end{equation}
where $\dot M$ denotes the evaporation rate of the binary system. At leading order, this correction affects only the phase of the waveform.

Accordingly, the amplitude correction function vanishes,
\begin{equation}
\Xi_{\dot{ M}}=0,
\end{equation}
while the phase correction function depends on the symmetric mass ratio $\eta$ and can be written as
\begin{equation}
\Upsilon_{\dot{ M}}=\frac{25}{851968}\frac{3-26\eta+34\eta^2}{\eta^{2/5}(1-2\eta)}.
\end{equation}

As a result, all logarithmic derivatives of the correction functions with respect to the waveform parameters vanish, except for 
\begin{equation}
	\frac{\partial\ln\Upsilon_{\dot{M}}}{\partial\ln\eta}
=\eta \frac{-26+68\eta}{3-26\eta+34\eta^2}-\frac{2}{5}+\frac{2\eta}{1-2\eta}.
\end{equation}

\subsection{Dynamical Friction}
The presence of a surrounding medium, such as dark matter or gas, induces a dynamical friction force acting on the binary components and leads to an additional secular phase shift in the inspiral waveform. At leading order, this effect enters the gravitational-wave phase at the -5.5PN order. In the ppE framework adopted in this work, the dynamical friction effect is parameterized as
\begin{equation}
\delta=\rho,\qquad
b=-16,
\end{equation}
where $\rho$ characterizes the gas density. At leading order, this correction affects only the phase of the waveform.

Accordingly, the amplitude correction function vanishes,
\begin{equation}
\Xi_{\rm DF}=0,
\end{equation}
while the phase correction function depends on the symmetric mass ratio $\eta$ and can be written as
\begin{equation}
\Upsilon_{\rm DF}(f)=
-\frac{25\pi (3\eta-1) \mathcal M^2}{739328 \eta^2} 
\gamma_{\rm DF}(f),
\end{equation}
with
\begin{equation}
	\gamma_{\rm DF}(f)=
-247\ln \left(\frac{f}{f_{\rm DF}}\right)
-39
+304\ln 20
+38\ln \left(\frac{3125}{8}\right)
\end{equation}
\begin{equation}
	f_{\rm DF}=\frac{c_s}{22\pi(m_1+m_2)}
=\frac{c_s}{22\pi} \frac{\eta^{3/5}}{\mathcal M},
\end{equation}
where $c_s$ is the sound speed.

All derivatives of the correction functions can be derived and written as
\begin{equation}
	\frac{\partial\ln\Upsilon_{\rm DF}}{\partial\ln\mathcal M}
=2-\frac{247}{\gamma_{\rm DF}(f)},
\end{equation}
\begin{equation}
	\frac{\partial\ln\Upsilon_{\rm DF}}{\partial\ln\eta}
=\left(\frac{3\eta}{3\eta-1}-2\right)+\frac{741}{5 \gamma_{\rm DF}(f)},
\end{equation}
while all other terms are 0.

\bibliographystyle{spphys}
\bibliography{references}  

\end{document}